\documentclass[11pt]{article}
\usepackage{epsfig} 
\newcommand{\sect}[1]{ \section{#1} \setcounter{equation}{0} }

\newcommand{\pslash}{p \! \! \! /} 
\newcommand{\qslash}{q \! \! \! /}

\newcommand{\half}{\mbox{\small{$\frac{1}{2}$}}} 
\newcommand{\third}{\mbox{\small{$\frac{1}{3}$}}} 
 
\newcommand{\pitwo}{\mbox{\small{$\frac{\pi}{2}$}}} 
\newcommand{\pisix}{\mbox{\small{$\frac{\pi}{6}$}}} 
\newcommand{\MSbar}{\overline{\mbox{MS}}} 
\newcommand{\MSbars}{\overline{\mbox{\footnotesize{MS}}}} 
\newcommand{\alts}{\mbox{\footnotesize{alt}}} 
\newcommand{\Nf}{N_{\!f}}

\begin{document}
\title{RI${}^\prime$/SMOM scheme amplitudes for quark currents at two loops}
\author{J.A. Gracey, \\ Theoretical Physics Division, \\ 
Department of Mathematical Sciences, \\ University of Liverpool, \\ P.O. Box 
147, \\ Liverpool, \\ L69 3BX, \\ United Kingdom.} 
\date{} 
\maketitle 

\vspace{5cm} 
\noindent 
{\bf Abstract.} We determine the two loop corrections to the Green's function
of a quark current inserted in a quark $2$-point function at the symmetric
subtraction point. The amplitudes for the scalar, vector and tensor currents
are presented in both the $\MSbar$ and RI${}^\prime$/SMOM renormalization 
schemes. The RI${}^\prime$/SMOM scheme two loop renormalization for the scalar 
and tensor cases agree with previous work. The vector current renormalization
requires special treatment as it must be consistent with the Slavnov-Taylor 
identity which we demonstrate. We also discuss the possibility of an 
alternative definition of the RI${}^\prime$/SMOM scheme in the case of the 
tensor current. 

\vspace{-17cm}
\hspace{13.5cm}
{\bf LTH 906}

\newpage

\sect{Introduction.}

Non-abelian quantum field theory underlies the strong nuclear force which binds
quarks and gluons into hadrons. At high energy these quarks and gluons are
asymptotically free, \cite{1,2}, and so a good approximation to the physics of 
hadrons in deep inelastic scattering can be achieved by perturbation theory. 
When the coupling constant, $g$, is small then the only difficulty is the
actual computation of a large number of Feynman diagrams which prevents one 
from obtaining precise estimates. However, the physics of the specific 
structure of hadrons resides in the non-perturbative or low energy r\'{e}gime 
where, because the coupling constant is large, then perturbation theory is not 
applicable. Instead one focuses on the computation of matrix elements involving
the relevant operators for the hadrons or deep inelastic scattering process. In
principle such matrix elements can be measured accurately by using a lattice 
regularization of the non-abelian gauge theory. If one has access to powerful 
enough computers then one can build a solid picture of the dependence of the
matrix elements with momentum scales. One issue which arises in the lattice 
computations is that whilst concentrating on the low energy aspect, the 
resulting matrix elements must still match onto the high energy behaviour which
one can calculate in perturbation theory. This is not a trivial exercise. For 
instance, the operators one has to consider undergo renormalization. In 
perturbation theory the anomalous dimensions of key operators are known to at 
least three loops. See, for instance, \cite{3,4,5,6,7,8,9}. However, this is
invariably in the standard (non-physical) renormalization scheme known as 
$\MSbar$. In this scheme essentially only the basic infinities with respect to 
the regularizing parameter are subtracted leaving the finite parts unsubtracted
in the remaining part of the Green's function. Invariably one uses dimensional 
regularization in $d$~$=$~$4$~$-$~$2\epsilon$ dimensions. Therefore, the finite
parts of the matrix elements at high energy are a reflection of the scheme.

By contrast, the lattice uses various renormalization schemes which are 
different and physical. So to perform any matching in the overlap region 
requires knowledge of the matrix element in the {\em same} scheme whether this
is $\MSbar$ or a lattice based scheme. The latter set of schemes are chosen
primarily to reduce the financial cost of any numerical evaluation. For 
instance, derivatives within an operator or at any point require more
computation. So the suite of lattice schemes are designed to minimize such
complications. In earlier work the regularization invariant (RI) scheme and its
modified version (RI${}^\prime$) were defined in lattice computations, 
\cite{10,11}, and later developed to three and four loops for the continuum in 
several articles, both for the Landau gauge, \cite{12}, and general linear
covariant gauges, \cite{13}. Indeed matrix elements for deep inelastic
scattering operators were evaluated to three loops in RI${}^\prime$ in
\cite{13,14,15}. More recently a variation on the RI${}^\prime$ scheme has been
developed, \cite{16}. This is specifically related to matrix elements and
designed to overcome a problem with potential infrared singularities. In 
essence the RI${}^\prime$ scheme for $3$-point and higher Green's functions 
involves subtracting the divergences at an exceptional momentum configuration.
In other words the operator insertion is at zero momentum. To avoid this
exceptional point and hence the related infrared issue, the RI${}^\prime$/SMOM
renormalization scheme was introduced, \cite{16}. The annotation indicates that
there is non nullification of any of the external momenta in a $3$-point
function. Indeed the external momenta are non-zero and their squares are all
fixed to be at the same value when the Green's function is renormalized. Hence
one refers to it as a symmetric subtraction point. 

Initially this scheme was applied to the scalar, vector and tensor currents at 
one loop, \cite{16}, and to the scalar and tensor at two loops in \cite{17,18}.
More recently it has been extended to various low moment operators used in deep 
inelastic scattering at one loop, \cite{19}. However, the main focus of 
\cite{17,18} was the construction of the anomalous dimensions and thence the 
conversion functions from the RI${}^\prime$/SMOM scheme to the $\MSbar$ one. 
These are central to any mapping of lattice results to the perturbative region 
for measurement comparisons. However, when one undertakes any measurement on 
the lattice the Green's function with the operator insertion has free Lorentz
indices and therefore it has to be written in terms of a set of basis tensors. 
Then different combinations of the free Lorentz indices can be determined and 
information on the associated scalar amplitudes extracted. Whilst 
\cite{17,18} concentrated on the operator renormalization, the explicit values 
of the individual two loop scalar amplitudes were not given which would be 
invaluable for lattice measurements. Therefore, it is the main purpose of this 
article to provide that information at two loops not only for the Green's 
function with scalar and tensor current insertions to augment the work of 
\cite{17,18} but also for the vector current. This is partly because the latter
has not been treated within the RI${}^\prime$/SMOM formalism at {\em two} loops
but mainly because it underlies the renormalization of the deep inelastic 
scattering operators considered at one loop in \cite{19}. The amplitudes for 
those two sets of operators will be considered separately, \cite{20}. The case 
of the vector current is special as its renormalization is connected with the 
Slavnov-Taylor identity as discussed in \cite{16}. However, one feature of the 
tensor current is that the actual definition of the RI${}^\prime$/SMOM scheme 
for such operators is not unique. This is because there is a relatively large 
set of basis tensors due to the free Lorentz indices. Therefore, a different 
basis choice would lead to different scheme definitions as we will indicate. In
addition the way one projects out the part of the Green's function whose finite
part will be absorbed into the operator renormalization constant is also 
subject to a large degree of choice. As there is a range of ways of defining 
the RI${}^\prime$/SMOM scheme for the tensor current we will give one 
alternative for illustration but will also present all the amplitudes for the 
currents considered here in the $\MSbar$ scheme too. So an interested reader 
has the liberty to toy with variations on the RI${}^\prime$/SMOM scheme 
definition of \cite{16,17,18}. 

The article is organized as follows. General aspects of the computations we
perform as well as the techniques used to carry them out are given in section 
two. The three specific operators we consider and the details associated with
each are discussed in the three subsequent sections. Aspects of the conversion 
functions are given in section six including that for the alternative scheme 
devised for the tensor current to allow one to contrast with that of 
\cite{17,18}. Our conclusions are given in section seven. The main results are 
presented in a series of Tables.  

\sect{Preliminaries.}

To start with we focus on the generalities of the computation we are interested
in. The three basic quark currents are the scalar, vector and tensor currents
and we use the compact notation introduced in \cite{19,21} 
\begin{equation}
S ~ \equiv ~ \bar{\psi} \psi ~~~,~~~ 
V ~ \equiv ~ \bar{\psi} \gamma^\mu \psi ~~~,~~~
T ~ \equiv ~ \bar{\psi} \sigma^{\mu\nu} \psi 
\end{equation}
where $\sigma^{\mu\nu}$~$=$~$\half [\gamma^\mu , \gamma^\nu]$ is antisymmetric.
Each of these operators, ${\cal O}$, is inserted into a quark $2$-point Green's
function where the two independent external momenta, $p$ and $q$, flow in
through each external quark leg as illustrated in Figure $1$. As we will be
concentrating on the renormalization of the operators and the consequent finite
parts of the Green's function of Figure $1$ at a symmetric renormalization
point, we note that from here on we take, \cite{16,17,18},
\begin{equation}
p^2 ~=~ q^2 ~=~ ( p + q )^2 ~=~ -~ \mu^2 
\end{equation}
which implies
\begin{equation}
pq ~=~ \frac{1}{2} \mu^2 
\end{equation}
where $\mu$ is the mass scale associated with the renormalization point. At 
this point the Green's function can be written in terms of a set of scalar
amplitudes with respect to some basis of Lorentz tensors. Even though the 
scalar current has no free Lorentz indices there are two independent amplitudes
as the two independent momenta lead to two independent structures deriving from
the $\gamma$-matrices. This was discussed in \cite{19} but we note that when 
either of the external momenta is nullified, equating to the RI${}^\prime$
momentum configuration, then there is only one independent tensor in the basis.
Therefore, in general we can write the Green's function for each of our 
operators ${\cal O}^i_{\mu_1 \ldots \mu_n}$ separately as 
\begin{equation}
\left. \left\langle \psi(p) {\cal O}^i_{\mu_1 \ldots \mu_{n_i}}(-p-q) 
\bar{\psi}(q) \right\rangle \right|_{p^2 = q^2 = - \mu^2} ~=~ \sum_{k=1}^{n_i} 
{\cal P}^i_{(k) \, \mu_1 \ldots \mu_{n_i} }(p,q) \, 
\Sigma^{{\cal O}^i}_{(k)}(p,q) 
\end{equation}
where $i$ is the label corresponding to the operator, which is either $S$, $V$
or $T$ here. The scalar amplitudes are denoted by 
$\Sigma^{{\cal O}^i}_{(k)}(p,q)$ with ${\cal P}^i_{(k) \, \mu_1 \ldots 
\mu_{n_i} }(p,q)$ corresponding to the basis tensors. The latter were given in
\cite{19} and were derived by writing all the one loop Feynman diagrams in 
terms of a basic set of master tensor integrals. In other words each diagram 
was stripped of all features to leave purely basic integrals. These were 
evaluated by standard methods and then substituted back so that the computation
could be completed by contracting the Lorentz indices with the stripped off 
$\gamma$-matrices. The reason for proceeding in this way was to ensure that no 
tensors contributing to the basis in the decompostion were omitted. Ordinarily 
one would construct the basis by building all possible tensors from the basic 
tensors such as $\eta_{\mu\nu}$, $\gamma$-matrices and independent momenta. 
Although here there are two momenta it turns out that even for simple currents 
the basis can be quite large. This is aside from the fact that we are only 
constructing the basis at the symmetric subtraction point where the values of 
$p^2$, $q^2$ and $(p+q)^2$ are all equal. Away from this point the basis of 
tensors will be significantly larger but we are not interested in the form of 
the amplitudes not at the symmetric point. 

\begin{figure}[ht]
\hspace{6cm}
\epsfig{file=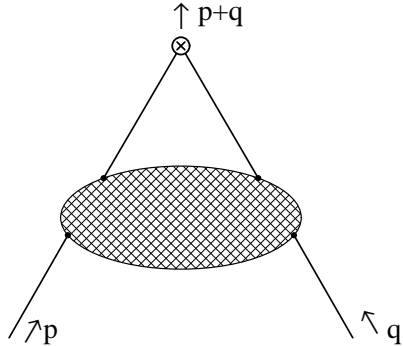,height=5cm}
\vspace{0.5cm}
\caption{Momentum flow for the Green's function, $\left\langle \psi(p)
{\cal O}^i_{\mu_1 \ldots \mu_{n_i}} (-p-q) \bar{\psi}(q) \right\rangle$.}
\end{figure}

As the scalar amplitudes are the quantities we seek then it is possible to
compute each individually via a projection onto the Green's function itself. In
other words 
\begin{equation}
\Sigma^{{\cal O}^i}_{(k)}(p,q) ~=~ {\cal M}^i_{kl} 
{\cal P}^{i ~\, \mu_1 \ldots \mu_{n_i}}_{(l)}(p,q) \left. \left(  
\left\langle \psi(p) {\cal O}^i_{\mu_1 \ldots \mu_{n_i}}(-p-q) \bar{\psi}(q) 
\right\rangle \right) \right|_{p^2=q^2=-\mu^2} 
\end{equation}
where ${\cal M}^i_{kl}$ is a matrix of rational polynomials in $d$ with $k$ and
$l$ labelling the different amplitudes relative to the ordering of the tensors 
in the basis, \cite{19}. Thoughout we use dimensional regularization in 
$d$~$=$~$4$~$-$~$2\epsilon$ dimensions. This matrix is computed from the actual
tensors themselves. First, if we define the matrix  
\begin{equation}
{\cal N}^i_{kl} ~=~ \left. {\cal P}^i_{(k) \, \mu_1 \ldots \mu_{n_i}}(p,q)
{\cal P}^{i ~\, \mu_1 \ldots \mu_{n_i}}_{(l)}(p,q) \right|_{p^2=q^2=-\mu^2} 
\end{equation}
then ${\cal M}^i_{kl}$ is its inverse. Of course the choice of basis tensors is
not unique and we use the one constructed for each operator at one loop. Whilst
that involved stripping the spinor structure off all the Feynman graphs and
then substituting the master tensor integrals, such an exercise would be too
difficult and cumbersome to implement at two loops. Therefore, we use the 
projection of the Green's function as outlined here. This was also used at one
loop and reproduced the direct evaluation results of \cite{19}. Therefore, we 
are confident that we have a complete basis. Though there are additional checks
on the results such as agreement with the Slavnov-Taylor identity, in the case 
of the vector, as will be discussed later.  

With our general decomposition of the Green's function we can now discuss the
renormalization and the method to define the RI${}^\prime$/SMOM scheme 
renormalization constants. Of the set of scalar amplitudes, one or more will
contain the poles in $\epsilon$. If we denote this amplitude, or set of
amplitudes, by the label $0$ then the renormalization constant for the operator
is defined by the condition, \cite{16,17,18},  
\begin{equation}
\left. \lim_{\epsilon\rightarrow 0} \left[ 
Z^{\mbox{\footnotesize{RI$^\prime$}}}_\psi 
Z^{\mbox{\footnotesize{RI$^\prime$/SMOM}}}_{\cal O} \Sigma^{\cal O}_{(0)}(p,q) 
\right] \right|_{p^2 \, = \, q^2 \, = \, - \mu^2} ~=~ 1 ~. 
\label{rencond}
\end{equation} 
In other words there are no $O(a)$ corrections after renormalization where we
set $a$~$=$~$g^2/(16\pi^2)$ and $g$ is the coupling constant. In addition we 
note that the origin of the second aspect of this scheme definition is that the
quark wave function renormalization is carried out in the RI${}^\prime$ scheme.
Briefly, this scheme is defined by ensuring that after renormalization there 
are no $O(a)$ corrections in $2$-point functions but for $3$-point functions 
and higher the renormalization is performed as one does in the $\MSbar$ scheme.
This RI${}^\prime$ scheme and its sister scheme, RI, were introduced in 
\cite{10,11} and examined in the continuum case at three and four loops, 
\cite{12,13,14}. The latter work was for a linear covariant gauge fixing. Such 
computations are important when one is converting from schemes such as 
RI${}^\prime$ or RI${}^\prime$/SMOM to $\MSbar$ as one has to express the 
parameters of each scheme in terms of the same parameters of the other scheme. 
It transpires that the coupling constants are in direct correspondence,
\cite{13},  
\begin{equation}
a_{\mbox{\footnotesize{RI$^\prime$}}} ~=~
a_{\mbox{\footnotesize{$\MSbar$}}} ~+~ O \left(
a_{\mbox{\footnotesize{$\MSbar$}}}^5 \right)
\end{equation}
to the order they were computed where we indicate the scheme to which the
variables relate via the subscript. However, for a linear covariant gauge the
gauge parameter, $\alpha$, is not in one-to-one correspondence except in the
Landau gauge as, \cite{13}, 
\begin{eqnarray}
\alpha_{\mbox{\footnotesize{RI$^\prime$}}}
&=& \left[ 1 + \left( \left( - 9 \alpha_{\mbox{\footnotesize{$\MSbar$}}}^2
- 18 \alpha_{\mbox{\footnotesize{$\MSbar$}}} - 97 \right) C_A + 80 T_F \Nf
\right) \frac{a_{\mbox{\footnotesize{$\MSbar$}}}}{36} \right. \nonumber \\
&& \left. ~+~ \left( \left( 18 \alpha_{\mbox{\footnotesize{$\MSbar$}}}^4
- 18 \alpha_{\mbox{\footnotesize{$\MSbar$}}}^3
+ 190 \alpha_{\mbox{\footnotesize{$\MSbar$}}}^2
- 576 \zeta(3) \alpha_{\mbox{\footnotesize{$\MSbar$}}}
+ 463 \alpha_{\mbox{\footnotesize{$\MSbar$}}} + 864 \zeta(3) - 7143 \right)
C_A^2 \right. \right. \nonumber \\
&& \left. \left. ~~~~~~~+~ \left( -~ 320
\alpha_{\mbox{\footnotesize{$\MSbar$}}}^2
- 320 \alpha_{\mbox{\footnotesize{$\MSbar$}}} + 2304 \zeta(3)
+ 4248 \right) C_A T_F \Nf \right. \right. \nonumber \\
&& \left. \left. ~~~~~~~+~ \left( -~ 4608 \zeta(3) + 5280 \right) C_F T_F \Nf
\right) \frac{a^2_{\mbox{\footnotesize{$\MSbar$}}}}{288} \right]
\alpha_{\mbox{\footnotesize{$\MSbar$}}} ~+~ O \left(
a^3_{\mbox{\footnotesize{$\MSbar$}}} \right) 
\label{almap}
\end{eqnarray}
to two loops. The full three loop result is given in \cite{13}. These relations 
between the parameters of each scheme are required when converting the
amplitudes. The group Casimirs are defined in the usual way by
\begin{equation}
T^a T^a ~=~ C_F ~~~,~~~ f^{acd} f^{bcd} ~=~ C_A \delta^{ab} ~~~,~~~
\mbox{Tr} \left( T^a T^b \right) ~=~ T_F \delta^{ab}
\end{equation}
where $T^a$ are the generators of the colour group whose structure functions 
are $f^{abc}$. Throughout we work with $\Nf$ massless quarks so that all our
expressions for amplitudes are for the chiral limit. In principle quark masses
could be included. However, the basic master scalar Feynman integrals at two
loops with massive propagators are not known exactly for the symmetric point.
So quark mass dependence for the scalar amplitudes of the Green's function we
are computing could only be determined in, say, the small quark mass limit. In
connection with the relation of the parameters between schemes we should note
that in \cite{16} a different parameter mapping was chosen. There it was 
assumed that the gauge parameter was the same in both schemes in the same way
that the coupling constants are. Here we will present all our results for the 
non-Landau schemes in what we regard as the full RI${}^\prime$ context which
led to (\ref{almap}). 

The next aspect of (\ref{rencond}) which we need to draw attention to is the
way in which the operator renormalization constant is actually defined. For
instance, in writing (\ref{rencond}) we are making a basis dependent statement.
The choice of the basis tensors is purely arbitrary and another choice with the
same underlying criterion of having no $O(a)$ correction will clearly lead to a
different numerical value of the corrections in the renormalization constant
itself. Moreover, the amplitudes will also have the same degree of 
arbitrariness. This is not the same as scheme dependence. In that situation if
one could compute the amplitudes to all orders the physics would not be 
affected by the renormalization scheme choice. However, as one has to truncate
series in quantum field theory due to the limit of calculability it is clear
that the construction of conversion functions, such as those of 
\cite{16,17,18}, could be affected by the basis choice. Though in
\cite{16,17,18} the approach used was to project the Green's function with the 
appropriate Born term before rendering that projection to have no $O(a)$ part. 
This defined the RI${}^\prime$/SMOM scheme for the scalar and tensor currents. 
However, once the full scalar amplitudes have been computed it is clear that 
there is in fact a sizeable number of ways of defining a so called 
RI${}^\prime$/SMOM scheme. This was discussed in the one loop context in
\cite{19} where it is becomes a more important issue for the operators used in 
deep inelastic scattering. Therefore, we will discuss a possible variation on 
the scheme of \cite{17,18} for the tensor current at two loops. In addition, 
partly because of these general aspects we will provide all the amplitudes in 
the reference $\MSbar$ scheme. This is primarily for lattice computations where
it is easier to convert to $\MSbar$ on the lattice before looking at the 
matching onto the ultraviolet part of the Green's function where perturbation 
theory is valid.  

Next we address some of the technical aspects of the computation. One feature
of the basis of tensors is that they involve products of $\gamma$-matrices.
This is because one can have contractions such as $\pslash$ and $\qslash$.
Therefore, we have chosen to work with the generalized $\Gamma$-matrices of
\cite{22,23,24} which are denoted by $\Gamma_{(n)}^{\mu_1 \ldots \mu_n}$. They
are defined by 
\begin{equation}
\Gamma_{(n)}^{\mu_1 \ldots \mu_n} ~=~ \gamma^{[\mu_1} \ldots \gamma^{\mu_n]}
\end{equation}
which is totally antisymmetric in all the Lorentz indices and the notation
includes the overall factor of $1/n!$. So, for instance,
$\sigma^{\mu\nu}$~$=$~$\Gamma_{(2)}^{\mu\nu}$. General properties are given in
\cite{25,26}. One advantage of this choice for the tensor basis is that in 
$\Gamma$-space 
\begin{equation}
\mbox{tr} \left( \Gamma_{(m)}^{\mu_1 \ldots \mu_m}
\Gamma_{(n)}^{\nu_1 \ldots \nu_n} \right) ~ \propto ~ \delta_{mn}
I^{\mu_1 \ldots \mu_m \nu_1 \ldots \nu_n} 
\label{gamtr}
\end{equation}
where $I^{\mu_1 \ldots \mu_m \nu_1 \ldots \nu_n}$ is the unit matrix in this
space. So there is a natural partition for the basis. The main tool to handle
the tedious algebra for manipulating tensor projections is the symbolic
manipulation language {\sc Form}, \cite{27}. The Feynman diagrams are generated
with the {\sc Qgraf} package, \cite{28}, and converted into {\sc Form} notation
whereby the colour and Lorentz indices are appended. For the Green's function
we are interested in with quark current insertions there are $1$ one loop graph
and $13$ two loop graphs to be computed. These graphs are readily broken up 
into a set of tensor integrals where the external momenta swamp any free 
Lorentz index when we have multiplied by the appropriate tensor of the 
projection basis. As these integrals are in essence $3$-point functions 
evaluated at the symmetric point the next stage is to break them down into the
known scalar master integrals. These are collected in \cite{16} but were 
derived in various articles, \cite{29,30,31,32}. The engine room of this aspect
of the computation is the Laporta algorithm, \cite{33}. Briefly, that method
allows one to build a redundant set of equations where basic Feynman graphs 
with irreducible numerators are all related by integration by parts and Lorentz
identities. These can then be solved as a tower of equations with the master 
integrals being the foundation for the many sets of integrals in predefined 
sectors. There are a variety of computer packages available to build a Laporta 
system, \cite{34,35}. However, we have used {\sc Reduze}, \cite{36}, which uses 
{\sc Ginac}, \cite{37}, and involves C$++$ at its root. For the particular 
Green's function we consider we need only build two topologies using the 
{\sc Reduze} package since all the Feynman graphs are either based on a two 
loop ladder or the non-planar ladder. Once the system is built it can be 
converted into {\sc Form} language and all the two loop tensor integrals for 
all the Feynman graphs at the symmetric subtraction point are written in terms 
of the known scalar master integrals listed in \cite{18}. A check on the 
{\sc Reduze} results is that we do reproduce the expressions given in 
\cite{16,17,18}. Though the expressions we give at two loops for the amplitudes
are new. 

As was evident in \cite{17,18} the final forms for the two loop anomalous
dimensions and associated conversion functions were surprisingly long. As we 
will be presenting a large number of amplitudes for the three currents in two 
renormalization schemes in order to save space we will collect the main results
in Tables\footnote{Attached to this article is an electronic file where all the
expressions presented in the Tables are available in a useable format.}. To do 
this we have had to split the amplitudes by their colour group structure and so
we have defined 
\begin{eqnarray}
\Sigma^{{\cal O}^i}_{(i)}(p,q) &=& \left( \sum_n c^{{\cal O}^i,(1)}_{(i)\,n} 
a^{(1)}_n \right) C_F a ~+~ 
\left( \sum_n c^{{\cal O}^i,(21)}_{(i)\,n} a^{(21)}_n \right) C_F T_F \Nf a^2 
\nonumber \\
&& +~ \left( \sum_n c^{{\cal O}^i,(22)}_{(i)\,n} a^{(22)}_n \right) 
C_F C_A a^2 ~+~ \left( \sum_n c^{{\cal O}^i,(23)}_{(i)\,n} a^{(23)}_n \right) 
C_F^2 a^2 ~+~ O(a^3) 
\label{notat}
\end{eqnarray}
where the parameters are in either scheme. Here we will adapt the convention 
that all expressions are in the RI${}^\prime$/SMOM scheme unless explicitly 
indicated to be in the $\MSbar$ scheme. The coupling constants are the same but
the gauge parameters are not, (\ref{almap}). In (\ref{notat}) the entities 
$a_n^{(k)}$ denote the basis of numbers which appear in that specific part of 
anomalous dimension. This includes, for instance, the gauge parameter
dependence and it is evident from the left hand column of each table what these
numbers actually are for each colour structure. In essence they relate to the 
form of the scalar master integrals at the symmetric subtraction point. The 
other entities, $c^{{\cal O}^i,(k)}_{(i)\,n}$, are the actual coefficients of 
the number basis. The summation label $n$ corresponds to the row of each table.
The superscript $k$ annotates the loop order, as the first number, and at two 
loop the second number references the colour group structure. Although the 
exact expressions at two loop represent all the information on the scalar 
amplitudes which we seek, we have relegated the Tables to the end of the 
article as it is the numerical evaluation which is ultimately of practical use. 
These expressions will be provided for each operator in succession in the next
few sections. Throughout the one loop amplitudes are in agreement with those of
\cite{16}. 

\sect{Scalar current.}

In this section we concentrate on the scalar current. The aim for this and the
other currents is to provide not only the anomalous dimensions in the full 
RI${}^\prime$/SMOM scheme but also the finite parts of all the amplitudes. 
Previously in \cite{17,18} the two loop anomalous dimension and conversion 
function were given. However, in order to assist with the extraction of results
from the lattice, measurements have to be made in different components of the 
Lorentz basis of tensors and therefore all the finite parts of the Green's 
functions need to be known accurately. As the amplitudes for the scalar current
have not been given at two loops we briefly report on this situation in this
section first. The decomposition into the projection tensors involves two 
tensors which are
\begin{equation}
{\cal P}^{S}_{(1)}(p,q) ~=~ \Gamma_{(0)} ~~~,~~~
{\cal P}^{S}_{(2)}(p,q) ~=~ \frac{1}{\mu^2} \Gamma_{(2)}^{pq}
\end{equation}
where we use the convention that if a momentum is contracted with a Lorentz
index of $\Gamma_{(n)}^{\mu_1 \ldots \mu_n}$ then we replace that index by the 
momentum to save space. The matrix used for constructing the explicit 
projection is given by, \cite{19},  
\begin{equation}
{\cal M}^S ~=~ \frac{1}{12} \left(
\begin{array}{cc}
3 & 0 \\
0 & - 4 \\
\end{array}
\right) ~.
\end{equation}
This is a simple example of the partitioning of the matrix as a consequence of
the choice of the $\Gamma_{(n)}$ basis. As noted previously to record the full 
two loop expressions for the amplitudes would be demanding on space. However, 
as there are only two amplitudes for the scalar case we present the results for
both schemes in Tables $1$ to $4$. The notation of (\ref{notat}) is used with 
the convention that the $\MSbar$ scheme results are annotated explicitly. 
Otherwise the results are in the RI${}^\prime$/SMOM scheme. For the scalar 
current the RI${}^\prime$/SMOM scheme renormalization condition is to project 
the Green's function with the Born term and then define the operator
renormalization constant so that there is no $O(a)$ finite part, 
\cite{16,17,18}. Since we are using the generalized $\Gamma$-matrix basis then 
for the scalar case this equates to ensuring that there are no $O(a)$ 
corrections to the channel $1$ amplitude. This is because of (\ref{gamtr}). 
Consequently there are no RI${}^\prime$/SMOM scheme coefficients for channel 
$1$ in these Tables. 

Throughout we use the standard basis for the numbers which appear in this 
symmetric subtraction point momentum configuration for this Green's function, 
\cite{17}. There $\psi(z)$ is the derivative of the logarithm of the Euler 
Gamma function,
\begin{equation}
s_n(z) ~=~ \frac{1}{\sqrt{3}} \Im \left[ \mbox{Li}_n \left( 
\frac{e^{iz}}{\sqrt{3}} \right) \right]
\end{equation}
where $\mbox{Li}_n(z)$ is the polylogarithm function, $\zeta(z)$ is the 
Riemann zeta function and $\Sigma$ is given by a combination of various
harmonic polylogarithms, \cite{17,32}, 
\begin{equation}
\Sigma ~=~ {\cal H}^{(2)}_{31} ~+~ {\cal H}^{(2)}_{43} ~.
\end{equation}
Whilst the exact two loop expressions are the output from the {\sc Form}
computation, for practical purposes expressing the result in numerical form 
will be more pragmatic for users. Therefore, we record this in an explicit 
equation for the case of $SU(3)$ not only for the scalar operator but also for 
the other cases we consider later. For the $\MSbar$ scheme we have  
\begin{eqnarray}
\left. \Sigma^{S}_{(1)}(p,q) \right|_{\MSbars} &=& -~ 1 ~-~ \left[ 
1.1040618 \alpha + 0.6455188 \right] a \nonumber \\
&& -~ \left[ 48.4885881 + 7.58534654 \alpha + 3.67844381 \alpha^2 
- 6.3468728 \Nf \right] a^2 ~+~ O(a^3) \nonumber \\
\left. \Sigma^{S}_{(2)}(p,q) \right|_{\MSbars} &=& \left[ 1.0417366 \alpha 
- 1.0417366 \right] a \nonumber \\
&& -~ \left[ 11.1668053 - 7.37922016 \alpha - 3.5592665 \alpha^2
- 0.4629940 \Nf \right] a^2 \nonumber \\
&& +~ O(a^3) ~.
\end{eqnarray}
Those for the RI${}^\prime$/SMOM scheme are 
\begin{eqnarray}
\Sigma^{S}_{(1)}(p,q) &=& -~ 1 ~+~ O(a^3) \nonumber \\
\Sigma^{S}_{(2)}(p,q) &=& \left[ 1.0417366 \alpha - 1.0417366 \right] a 
\nonumber \\
&& -~ \left[ 10.4943447 - 16.2776049 \alpha - 3.97172981 \alpha^2
- 0.7813024 \alpha^3 \right. \nonumber \\
&& \left. ~~~~+~ \left[ 1.1574851 \alpha - 0.4629940 \right] \Nf \right] a^2 
~+~ O(a^3) ~.
\end{eqnarray}
Throughout we use the numerical values
\begin{eqnarray}
\zeta(3) &=& 1.20205690 ~~,~~ \Sigma ~=~ 6.34517334 ~~,~~ 
\psi^\prime\left( \frac{1}{3} \right) ~=~ 10.09559713 \nonumber \\
\psi^{\prime\prime\prime}\left( \frac{1}{3} \right) &=& 488.1838167 ~~,~~ 
s_2\left( \frac{\pi}{2} \right) ~=~ 0.32225882 ~~,~~ 
s_2\left( \frac{\pi}{6} \right) ~=~ 0.22459602 \nonumber \\
s_3\left( \frac{\pi}{2} \right) &=& 0.32948320 ~~,~~ 
s_3\left( \frac{\pi}{6} \right) ~=~ 0.19259341 ~.
\end{eqnarray}
Clearly there is a weak dependence on the number of quarks in channel $2$. To 
reinforce an earlier point the gauge parameter $\alpha$ in these sets of
expressions is the variable in the RI${}^\prime$/SMOM scheme. Though the 
coupling constant, $a$, is in the same scheme it is an exact map of the 
$\MSbar$ variable. Moreover, as with other RI${}^\prime$/SMOM scheme anomalous 
dimensions of gauge invariant operators the higher loop corrections will depend
on the gauge parameter. This is because the scheme is a mass dependent one and 
not a mass independent one like $\MSbar$. Finally, we note that using 
(\ref{almap}) the RI${}^\prime$/SMOM scheme anomalous dimension is
\begin{eqnarray}
\left. \frac{}{} \gamma^S(a,\alpha) 
\right|_{\mbox{\footnotesize{RI$^\prime$/SMOM}}} &=& -~ 3 C_F a \nonumber \\
&& +~ \left[ \left[ ( 3 \alpha^2 + 9 \alpha + 66 ) \psi^\prime(\third)
- ( 2 \alpha^2 + 6 \alpha + 44 ) \pi^2 \right. \right. \nonumber \\
&& \left. \left. ~~~~~-~ 9 \alpha^2 - 27 \alpha - 555 \right] C_A ~-~
27 C_F \right. \nonumber \\
&& \left. ~~~~+~ \left[ 16 \pi^2 + 156 - 24 \psi^\prime(\third) \right]
T_F \Nf \right] \frac{C_F a^2}{18} ~+~ O(a^3) 
\end{eqnarray}
which agrees with \cite{16,17,18} when $\alpha$~$=$~$0$ where we will annotate
the anomalous dimensions with the scheme. The $\alpha$ dependent terms differ 
because of the different ways the renormalization of $\alpha$ is performed. We 
have chosen to define the $\alpha$ renormalization using the RI${}^\prime$
scheme of \cite{13} whereas in \cite{17} the $\alpha$ renormalization is taken 
to be in the $\MSbar$ scheme. As ultimately lattice computations are in the 
Landau gauge this difference in definitions would only be important in 
non-Landau linear covariant gauges.

\sect{Vector current.}

As the vector operator has not received attention at two loops, we devote this 
section to it in detail using the general notation discussed previously. First,
the basis of tensors used to decompose the Green's function at the symmetric 
subtraction point is, \cite{19}, 
\begin{eqnarray}
{\cal P}^{V}_{(1) \mu }(p,q) &=& \gamma_\mu ~~~,~~~
{\cal P}^{V}_{(2) \mu }(p,q) ~=~ \frac{{p}^\mu \pslash}{\mu^2} ~~~,~~~
{\cal P}^{V}_{(3) \mu }(p,q) ~=~ \frac{{p}_\mu \qslash}{\mu^2} ~, \nonumber \\
{\cal P}^{V}_{(4) \mu }(p,q) &=& \frac{{q}_\mu \pslash}{\mu^2} ~~~,~~~
{\cal P}^{V}_{(5) \mu }(p,q) ~=~ \frac{{q}_\mu \qslash}{\mu^2} ~~~,~~~
{\cal P}^{V}_{(6) \mu }(p,q) ~=~ \frac{1}{\mu^2} \Gamma_{(3) \, \mu p q} ~,
\end{eqnarray}
where the matrix used to perform the projection into the various amplitudes is,
\cite{19}, 
\begin{equation}
{\cal M}^V ~=~ \frac{1}{36(d-2)} \left(
\begin{array}{cccccc}
9 & 12 & 6 & 6 & 12 & 0 \\
12 & 16 (d - 1) &  8 (d - 1) &  8 (d - 1) & 4 (d + 2) & 0 \\
6 & 8 (d - 1) & 4 (4 d - 7) &  4 (d - 1) & 8 (d - 1) & 0 \\
6 & 8 (d - 1) &  4 (d - 1) & 4 (4 d - 7) & 8 (d - 1) & 0 \\
12 & 4 (d + 2) &  8 (d - 1) &  8 (d - 1) & 16 (d - 1) & 0 \\
0 & 0 & 0 & 0 & 0 & - 12 \\
\end{array}
\right) ~.
\end{equation}
With this setup we have applied the computational algorithm to extract the
RI${}^\prime$/SMOM renormalization constant. To do this for the 
RI${}^\prime$/SMOM scheme requires a different approach compared to the other quark
currents. The reason for this is that the vector current is a physical operator
and therefore its renormalization is already determined by general 
considerations. Specifically as it is physical its anomalous dimension is zero
which is widely known in the $\MSbar$ scheme. However, once it is accepted that
the renormalization constant is unity in one scheme then it is unity in all
other schemes, \cite{38}. Underlying this is the Slavnov-Taylor identity which
relates the renormalization of the Green's function with the divergence of the
operator to the renormalization of the quark $2$-point functions. Indeed this 
was discussed in \cite{13} for the RI${}^\prime$ scheme and demonstrated to be 
consistent to three loops. The situation for the RI${}^\prime$/SMOM computation
is more involved as there are two momenta flowing through the Green's function 
with the operator insertion. Therefore, to reproduce the Slavnov-Taylor 
identity the Green's function of Figure $1$ has to be contracted with the
vector $(p+q)_\mu$. This is one reason why we have to decompose the Green's
function into a basis of projection tensors. Once we have established this then
the contraction can proceed. For the case of the RI${}^\prime$ scheme this 
aspect of the reconciliation with the Slavnov-Taylor identity is simplified 
significantly because there is only one momentum flowing through the Green's 
function. Indeed put another way the contraction of the graph of Figure $1$ 
with $(p+q)_\mu$ will effectively become the renormalization condition for
$Z^V$ but will naturally produce unity as expected from general theorems. 
Whilst we are focusing on this feature for the vector current in detail here it
transpires that the same issue arises for operators which are used in deep 
inelastic scattering. This was discussed at one loop in \cite{19} for the sets 
of operators labelled $W_2$ and $W_3$ in the notation of \cite{21}. This is 
because the contraction of Lorentz indices of one of the operators in each set 
is equivalent to the divergence of the vector current. Therefore, the 
renormalization of those operators also has to be consistent with the 
Slavnov-Taylor identity.
 
To focus on the issue we concentrate on the $\MSbar$ renormalization first. The
explicit results for each of the amplitudes is given in Tables $5$ to $8$. The 
numerical values for the amplitudes for $SU(3)$ are
\begin{eqnarray}
\left. \Sigma^{V}_{(1)}(p,q) \right|_{\MSbars} &=& -~ 1 ~+~ \left[ 1.6249301
- 0.5831936 \alpha \right] a \nonumber \\
&& +~ \left[ 6.1248321 - 3.2229010 \alpha - 1.8119992 \alpha^2 + 0.2362586 \Nf
\right] a^2 ~+~ O(a^3) \nonumber \\
\left. \Sigma^{V}_{(2)}(p,q) \right|_{\MSbars} &=&
\left. \Sigma^{V}_{(5)}(p,q) \right|_{\MSbars} \nonumber \\
&=& \left[ 1.4720824 + 0.3056953 \alpha \right] a \nonumber \\
&& +~ \left[ 18.7974908 + 4.5957818 \alpha + 1.0954082 \alpha^2 
- 1.3299518 \Nf \right] a^2 ~+~ O(a^3) \nonumber \\
\left. \Sigma^{V}_{(3)}(p,q) \right|_{\MSbars} &=& 
\left. \Sigma^{V}_{(4)}(p,q) \right|_{\MSbars} \nonumber \\
&=& \left[ 1.7777778 + 1.1945842 \alpha \right] a \nonumber \\
&& +~ \left[ 44.3805855 + 12.1090504 \alpha + 4.2805934 \alpha^2 
- 2.8641975 \Nf \right] a^2 ~+~ O(a^3) \nonumber \\
\left. \Sigma^{V}_{(6)}(p,q) \right|_{\MSbars} &=& -~ 2.0834731 a \nonumber \\
&& -~ \left[ 39.7873696 + 0.3484662 \alpha + 0.1736228 \alpha^3 - 3.0094611 \Nf 
\right] a^2 \nonumber \\
&& +~ O(a^3) ~. 
\end{eqnarray}
As noted at one loop in \cite{19} there is a degree of symmetry which is also
evident at two loops. This is primarily due to the way in which we chose our 
basis of Lorentz tensors which is of course not unique. The fact that various
channels pair off to two loops can be regarded as an internal check on both the
construction of the projection matrix and the symbolic manipulation code used 
to perform the computation. In the tables for the vector case we do not 
reproduce columns for the channels $4$ and $5$ because of the above equalities 
which hold exactly to two loops. In order to see that the Slavnov-Taylor 
identity is satisfied to two loops in the $\MSbar$ scheme we have computed that
combination of the amplitudes which correspond to the Green's function of the 
insertion of the divergence of the vector current. With the contraction with 
$(p+q)_\mu$ there will be two terms. One will involve $\pslash$ and the other 
$\qslash$. For the former the combination gives
\begin{eqnarray}
&& \left. \Sigma^{V}_{(1)}(p,q) \right|_{\MSbars} ~-~ 
\frac{1}{2} \left. \Sigma^{V}_{(2)}(p,q) \right|_{\MSbars} ~-~ \frac{1}{2}
\left. \Sigma^{V}_{(5)}(p,q) \right|_{\MSbars} \nonumber \\
&& =~ -~ 1 ~-~ \alpha C_F a ~+~ \left[ \left[ 3 \zeta(3) - \frac{41}{4} 
+ 3 \zeta(3) \alpha - \frac{13}{2} \alpha - \frac{9}{8} \alpha^2 \right] C_A 
+ \frac{5}{8} C_F + \frac{7}{2} T_F \Nf \right] C_F a^2 \nonumber \\
&& ~~~~+~ O(a^3) 
\label{vecstid2ms}
\end{eqnarray}
and that for the latter is a different combination of amplitudes but produces
the same result. The right hand side of (\ref{vecstid2ms}) is clearly the 
finite part of the quark $2$-point function after renormalizing in the $\MSbar$
scheme. Therefore, given the fact that the Green's function is symmetric under 
interchange of $p$ and $q$ then this shows that the Slavnov-Taylor identity is 
correctly embedded within our computation with the vector current having a 
renormalization constant of unity. 

The situation for the RI${}^\prime$/SMOM scheme amplitudes is completely 
parallel to that of $\MSbar$. We have given the amplitudes for this case in
Tables $9$ to $12$ where we have used 
\begin{equation}
Z^V ~=~ 1 ~+~ O(a^3)
\end{equation}
for the renormalization of the vector current. As a brief summary the $SU(3)$
numerical values are 
\begin{eqnarray}
\Sigma^{V}_{(1)}(p,q) &=& -~ 1 ~+~ \left[ 1.6249301 + 0.7501398 \alpha
\right] a \nonumber \\
&& +~ \left[ 31.5890381 + 12.2494724 \alpha + 2.8130241 \alpha^2 + 0.5626048
\alpha^3 \right. \nonumber \\
&& \left. ~~~~-~ \left[ 2.0970747 + 0.8334886 \alpha \right] \Nf \right] 
a^2 ~+~ O(a^3) \nonumber \\ 
\Sigma^{V}_{(2)}(p,q) &=& \Sigma^{V}_{(5)}(p,q) \nonumber \\
&=& \left[ 1.4720824 + 0.3056953 \alpha \right] a \nonumber \\
&& +~ \left[ 18.7974908 + 5.1040424 \alpha + 1.1463575 \alpha^2 
+ 0.2292715 \alpha^3 \right. \nonumber \\
&& \left. ~~~~-~ \left[ 1.3299518 + 0.3396615 \alpha \right] \Nf \right] 
a^2 ~+~ O(a^3) \nonumber \\
\Sigma^{V}_{(3)}(p,q) &=& \Sigma^{V}_{(4)}(p,q) \nonumber \\
&=& \left[ 1.7777778 + 1.1945842 \alpha \right] a \nonumber \\
&& +~ \left[ 44.3805855 + 19.3949024 \alpha + 4.4796908 \alpha^2 
+ 0.8959382 \alpha^3 \right. \nonumber \\
&& \left. ~~~~-~ \left[ 2.8641975 + 1.3273158 \alpha \right] \Nf \right] 
a^2 ~+~ O(a^3)
\nonumber \\ 
\Sigma^{V}_{(6)}(p,q) &=& -~ 2.0834731 a \nonumber \\
&& -~ \left[ 39.7873696 - 2.4294979 \alpha + 0.1736228 \alpha^2 - 3.0094611 \Nf 
\right] a^2 \nonumber \\
&& +~ O(a^3) ~. 
\end{eqnarray}
In order to see that the definition of $Z^V$ is consistent with the 
Slavnov-Taylor identity we have repeated the computation of (\ref{vecstid2ms})
for the RI${}^\prime$/SMOM amplitudes. In this case the piece involving
$\pslash$ produces 
\begin{equation}
\Sigma^{V}_{(1)}(p,q) ~-~ \frac{1}{2} \Sigma^{V}_{(2)}(p,q) ~-~ \frac{1}{2}
\Sigma^{V}_{(5)}(p,q) ~=~ -~ 1 ~+~ O(a^3) ~.
\label{vecstid2}
\end{equation}
Here there are no corrections which is consistent with the RI${}^\prime$/SMOM
scheme since, \cite{16}, it uses the RI${}^\prime$ quark wave function 
renormalization. This is chosen in such a way that the quark $2$-point function
has no $O(a)$ corrections. In other words the finite part of that $2$-point 
function is absorbed into the finite part of the wave function renormalization 
constant. Therefore, the vector current anomalous dimension 
\begin{equation}
\left. \frac{}{} \gamma^V(a,\alpha) 
\right|_{\mbox{\footnotesize{RI$^\prime$/SMOM}}} ~=~ O(a^3) 
\end{equation}
is consistent with the underlying Slavnov-Taylor identity. Whilst $\left. 
\frac{}{} \gamma^V(a,\alpha) \right|_{\mbox{\footnotesize{RI$^\prime$/SMOM}}}$
is zero to all orders we include the order symbol merely to record the order to
which we have explicitly verified this to. 

\sect{Tensor current.}

Next we record the parallel situation for the tensor operator noting only the
points where there are differences from earlier work and discussions. The 
Lorentz tensor basis for the amplitude decomposition is, \cite{19},  
\begin{eqnarray}
{\cal P}^{T}_{(1) \mu \nu }(p,q) &=& \Gamma_{(2) \, \mu\nu} ~~~,~~~
{\cal P}^{T}_{(2) \mu \nu }(p,q) ~=~ \frac{1}{\mu^2} \left[ p_\mu q_\nu - p_\nu
q_\mu \right] \Gamma_{(0)} ~, \nonumber \\
{\cal P}^{T}_{(3) \mu \nu }(p,q) &=& \frac{1}{\mu^2}
\left[ \Gamma_{(2) \, \mu p} p_\nu - \Gamma_{(2) \, \nu p} p_\mu
\right] ~~~,~~~
{\cal P}^{T}_{(4) \mu \nu }(p,q) ~=~ \frac{1}{\mu^2}
\left[ \Gamma_{(2) \, \mu p} q_\nu - \Gamma_{(2) \, \nu p} q_\mu
\right] ~, \nonumber \\
{\cal P}^{T}_{(5) \mu \nu }(p,q) &=& \frac{1}{\mu^2}
\left[ \Gamma_{(2) \, \mu q} p_\nu - \Gamma_{(2) \, \nu q} p_\mu
\right] ~~~,~~~
{\cal P}^{T}_{(6) \mu \nu }(p,q) ~=~ \frac{1}{\mu^2}
\left[ \Gamma_{(2) \, \mu q} q_\nu - \Gamma_{(2) \, \nu q} q_\mu
\right] ~, \nonumber \\
{\cal P}^{T}_{(7) \mu \nu }(p,q) &=& \frac{1}{\mu^4}
\left[ \Gamma_{(2) \, p q} p_\mu q_\nu - \Gamma_{(2) \, p q} p_\nu q_\mu
\right] ~~~,~~~
{\cal P}^{T}_{(8) \mu \nu }(p,q) ~=~ \frac{1}{\mu^2}
\Gamma_{(4) \, \mu \nu p q} ~.
\end{eqnarray}
The matrix to determine the explicit projection is defined as, \cite{19},  
\begin{equation}
{\cal M}^T ~=~ \frac{1}{36(d-2)(d-3)} \left(
\begin{array}{cc}
{\cal M}^T_{11} & {\cal M}^T_{12} \\
{\cal M}^T_{21} & {\cal M}^T_{22} \\
\end{array}
\right) 
\end{equation}
with the four submatrices given by 
\begin{eqnarray}
{\cal M}^T_{11} &=& \left(
\begin{array}{cccc}
- 9 & 0 & - 12 & - 6 \\
0 & 6 (d-2) (d-3) & 0 & 0 \\
- 12 & 0 & - 8 (d-1) & - 4 (d-1) \\
- 6 & 0 & - 4 (d-1) & - 4 (2d-5) \\
\end{array}
\right) ~, \nonumber \\
{\cal M}^T_{12} &=& \left(
\begin{array}{cccc}
- 6 & - 12 & - 12 & - 6 \\
0 & 0 & 0 & 0 \\
- 4 (d-1) & - 2 (d+5) & - 8 (d-1) & 0 \\
- 2 (d-1) & - 4 (d-1) & - 4 (d-1) & 0 \\
\end{array}
\right) ~, \nonumber \\
{\cal M}^T_{21} &=& \left(
\begin{array}{cccc}
- 6 & 0 & - 4 (d-1) & - 2 (d-1) \\
- 12 & 0 & - 2 (d+5) & - 4 (d-1) \\
- 12 & 0 & - 8 (d-1) & - 4 (d-1) \\
0 & 0 & 0 & 0 \\
\end{array}
\right) ~, \nonumber \\
{\cal M}^T_{22} &=& \left(
\begin{array}{cccc}
- 4 (2d-5) & - 4 (d-1) & - 4 (d-1) & 0 \\
- 4 (d-1) & - 8 (d-1)  & - 8 (d-1) & 0 \\
- 4 (d-1) & - 8 (d-1) & - 8 (d-1) (d-2) & 0 \\
0 & 0 & 0 & 12 \\
\end{array}
\right) ~.
\end{eqnarray}
This produces the two loop $\MSbar$ numerical values for the amplitudes 
\begin{eqnarray}
\left. \Sigma^{T}_{(1)}(p,q) \right|_{\MSbars} &=& -~ 1 ~+~ \left[ 0.0623253
- 0.0623253 \alpha \right] a \nonumber \\
&& +~ \left[ 17.0099539 + 0.6409422 \alpha + 0.0544455 \alpha^2 - 1.6001145 \Nf
\right] a^2 ~+~ O(a^3) \nonumber \\
\left. \Sigma^{T}_{(2)}(p,q) \right|_{\MSbars} &=& \left[ 3.1252097 
+ 1.0417366 \alpha \right] a \nonumber \\
&& +~ \left[ 76.2022091 + 10.8541168 \alpha + 3.9065121 \alpha^2 
- 5.5559283 \Nf \right] a^2 \nonumber \\
&& +~ O(a^3) \nonumber \\
\left. \Sigma^{T}_{(3)}(p,q) \right|_{\MSbars} &=& 
\left. \Sigma^{T}_{(6)}(p,q) \right|_{\MSbars} \nonumber \\
&=& \left[ 0.3056953 \alpha - 0.3056953 \right] a \nonumber \\
&& -~ \left[ 1.8195802 - 3.2067998 \alpha - 1.0954082 \alpha^2 
- 0.3396615 \Nf \right] a^2 ~+~ O(a^3) \nonumber \\
\left. \Sigma^{T}_{(4)}(p,q) \right|_{\MSbars} &=& 
\left. \Sigma^{T}_{(5)}(p,q) \right|_{\MSbars} \nonumber \\
&=& \left[ 0.1528477 \alpha - 0.1528477 \right] a \nonumber \\
&& -~ \left[ 0.9097901 - 1.6033999 \alpha - 0.5477041 \alpha^2 
- 0.1698307 \Nf \right] a^2 ~+~ O(a^3) \nonumber \\
\left. \Sigma^{T}_{(7)}(p,q) \right|_{\MSbars} &=& 
\left[ 0.6113907 \alpha - 0.6113907 \right] a \nonumber \\
&& -~ \left[ 3.6391605 - 6.4135995 \alpha - 2.1908165 \alpha^2 
- 0.6793229 \Nf \right] a^2 ~+~ O(a^3) \nonumber \\
\left. \Sigma^{T}_{(8)}(p,q) \right|_{\MSbars} &=& 
-~ \left[ 1.0417366 \alpha + 3.1252097 \right] a \nonumber \\
&& -~ \left[ 76.2022091 + 10.8541168 \alpha + 3.9065121 \alpha^2 
- 5.5559283 \Nf \right] a^2 \nonumber \\
&& +~ O(a^3) 
\end{eqnarray}
with the full expressions for each amplitude in the $\MSbar$ scheme recorded in
Tables $13$ to $16$. Unlike the vector case there is no Slavnov-Taylor identity
to be satisfied by the renormalization constant of the tensor current. For the 
$\MSbar$ renormalization the renormalization constant is already known and we 
have used that here. However, as noted in \cite{19} there is a variety of ways 
of defining the renormalization in the RI${}^\prime$/SMOM scheme case. In
\cite{16,17,18} the renormalization was performed by first projecting the 
Green's function with the tensor of the Born term. For the tensor current this 
is $\Gamma_{(2)}^{\mu\nu}$. The RI${}^\prime$/SMOM scheme renormalization is 
then defined so that there is no $O(a)$ correction in this projection. However, 
there is no reason to regard this as the unique way of defining the 
renormalization constant. Now that we have the complete decomposition into the 
tensor basis, one could define the renormalization so that instead the 
coefficient of $\Gamma_{(2)}^{\mu\nu}$ has no $O(a)$ piece after 
renormalization. This tensor channel contains the poles in $\epsilon$ which 
have to be removed. To us this is also a perfectly reasonable way to define the
scheme. Though it depends of course on the other elements of the tensor 
projection basis which is not unique. Indeed in \cite{19} this alternative way 
was studied and the one loop correction was found to be numerically smaller 
than that of \cite{16,17,18}. However it was not clear if this would persist to
next order. If not then it may be possible to improve the convergence by 
redefining the basis.

First, we record that we have followed the original RI${}^\prime$/SMOM scheme
definition of \cite{16,17,18} and reproduced precisely the full two loop 
renormalization constant and anomalous dimension of \cite{16,17,18}. This acts
as a non-trivial check on the {\sc Reduze} database of integrals we have 
constructed by the Laporta algorithm. However, the full RI${}^\prime$/SMOM 
amplitudes have not been presented and we record the numerical values are 
\begin{eqnarray}
\Sigma^{T}_{(1)}(p,q) &=& -~ 1 ~+~ \left[ 0.1528477 \alpha - 0.1528477 
\right] a \nonumber \\
&& -~ \left[ 0.9426788 - 2.9046959 \alpha - 0.7440869 \alpha^2 - 0.1146357
\alpha^3 \right. \nonumber \\ 
&& \left. ~~~~-~ \left[ 0.1698307 - 0.1698307 \alpha \right] \Nf 
\right] a^2 ~+~ O(a^3) \nonumber \\
\Sigma^{T}_{(2)}(p,q) &=& \left[ 1.0417366 \alpha + 3.1252097 \right] a 
\nonumber \\
&& +~ \left[ 76.8746697 + 18.8265135 \alpha + 5.2449633 \alpha^2 + 0.7813024
\alpha^3 \right. \nonumber \\ 
&& \left. ~~~~-~ \left[ 5.5559283 + 1.1574851 \alpha \right] \Nf 
\right] a^2 ~+~ O(a^3) \nonumber \\
\Sigma^{T}_{(3)}(p,q) &=& \Sigma^{T}_{(6)}(p,q) \nonumber \\
&=& \left[ 0.3056953 \alpha - 0.3056953 \right] a \nonumber \\
&& -~ \left[ 1.8853576 - 5.8093917 \alpha - 1.4881739 \alpha^2 - 0.2292715
\alpha^3 \right. \nonumber \\ 
&& \left. ~~~~-~ \left[ 0.3396615 - 0.3396615 \alpha \right] \Nf 
\right] a^2 ~+~ O(a^3) \nonumber \\
\Sigma^{T}_{(4)}(p,q) &=& \Sigma^{T}_{(5)}(p,q) \nonumber \\
&=& \left[ 0.1528477 \alpha - 0.1528477 \right] a \nonumber \\
&& -~ \left[ 0.9426788 - 2.9046959 \alpha - 0.7440869 \alpha^2 - 0.1146357
\alpha^3 \right. \nonumber \\ 
&& \left. ~~~~-~ \left[ 0.1698307 - 0.1698307 \alpha \right] \Nf 
\right] a^2 ~+~ O(a^3) \nonumber \\
\Sigma^{T}_{(7)}(p,q) &=& \left[ 0.6113907 \alpha - 0.6113907 \right] a 
\nonumber \\
&& -~ \left[ 3.7707152 - 11.6187834 \alpha - 2.9763478 \alpha^2 - 0.4585430
\alpha^3 \right. \nonumber \\ 
&& \left. ~~~~-~ \left[ 0.6793229 - 0.6793229 \alpha \right] \Nf 
\right] a^2 ~+~ O(a^3) \nonumber \\
\Sigma^{T}_{(8)}(p,q) &=& -~ \left[ 1.0417366 \alpha + 3.1252097 \right] a 
\nonumber \\
&& -~ \left[ 76.8746697 + 18.8265135 \alpha + 5.2449634 \alpha^2 + 0.7813024
\alpha^3 \right. \nonumber \\ 
&& \left. ~~~~-~ \left[ 5.5559283 + 1.1574851 \alpha \right] \Nf 
\right] a^2 ~+~ O(a^3) ~. 
\label{tensampl}
\end{eqnarray}
The full explicit results are given in Tables $17$ to $20$. Whilst the symmetry
derived from the interchange of the momenta $p$ and $q$ is evident numerically
here, this reflects the actual symmetry in the exact expressions which we have 
checked explicitly. So we have omitted those columns in the tables 
corresponding to the relations given in (\ref{tensampl}). To two loops the 
associated anomalous dimension is 
\begin{eqnarray}
\left. \frac{}{} \gamma^T(a,\alpha) 
\right|_{\mbox{\footnotesize{RI$^\prime$/SMOM}}} &=& C_F a \nonumber \\
&& +~ \left[ \left[ ( 9 \alpha^2 + 27 \alpha - 66 ) \psi^\prime(\third)
- ( 6 \alpha^2 + 18 \alpha - 44 ) \pi^2 \right. \right. \nonumber \\
&& \left. \left. ~~~~~-~ 9 \alpha^2 - 27 \alpha + 1035 \right] C_A ~-~
513 C_F \right. \nonumber \\
&& \left. ~~~~+~ \left[ 24 \psi^\prime(\third) - 16 \pi^2 - 252 \right]
T_F \Nf \right] \frac{C_F a^2}{54} ~+~ O(a^3) 
\label{tens2}
\end{eqnarray}
where $\left. \frac{}{} \gamma^T(a,0) 
\right|_{\mbox{\footnotesize{RI$^\prime$/SMOM}}}$ agrees with \cite{16,17,18}. 
As with the scalar current the difference with the $\alpha$ dependent terms
between (\ref{tens2}) and that of \cite{17} resides in the different ways the 
gauge parameter is renormalized. We again use that derived in the RI${}^\prime$
scheme, \cite{13}. There careful attention was given to ensuring that the full
three loop renormalization of QCD was consistent with the Slavnov-Taylor
identities in that scheme.

We close this section by discussing an alternative way of defining the tensor
current renormalization which was noted in \cite{19}. Instead of the approach 
of \cite{17} the finite part of associated with $\Gamma_{(2)}^{\mu\nu}$ in the 
tensor basis was used to define the renormalization constant, with respect to 
the basis we have introduced. This channel contains the singularities in 
$\epsilon$ which must always be absorbed in any scheme. Consequently we find 
the two loop anomalous dimension in this alternative scheme is
\begin{eqnarray}
\left. \frac{}{} \gamma^T(a,\alpha) 
\right|_{\mbox{\footnotesize{alt RI$^\prime$/SMOM}}} &=& C_F a \nonumber \\
&& +~ \left[ \left[ ( 45 \alpha^2 + 135 \alpha - 330 ) \psi^\prime(\third)
- ( 30 \alpha^2 + 90 \alpha - 220 ) \pi^2 \right. \right. \nonumber \\
&& \left. \left. ~~~~~-~ 81 \alpha^2 - 243 \alpha + 3501 \right] C_A ~-~
1539 C_F \right. \nonumber \\
&& \left. ~~~~+~ \left[ 120 \psi^\prime(\third) - 80 \pi^2 - 900 \right]
T_F \Nf \right] \frac{C_F a^2}{162} ~+~ O(a^3) ~. 
\label{alttens2}
\end{eqnarray}
For $SU(3)$ the numerical equivalent is 
\begin{eqnarray}
\left. \frac{}{} \gamma^T(a,\alpha) 
\right|_{\mbox{\footnotesize{alt RI$^\prime$/SMOM}}} &=& 1.3333333 a 
\nonumber \\
&& +~ \left[ 1.9065121 \alpha^2 + 5.7195362 \alpha + 40.9078004 -1.9674761 \Nf
\right] a^2 \nonumber \\
&& +~ O(a^3) ~.
\end{eqnarray}
Clearly the leading term is the same as the original RI${}^\prime$/SMOM scheme
as it ought to be. With this scheme choice the amplitudes are virtually the 
same as those for the original RI${}^\prime$/SMOM scheme of \cite{19} which we 
have presented already. The only differences are that the channel $1$
coefficients in Tables $17$ to $19$ are absent whilst the coefficients of 
Table $20$ are completely different. For this alternative scheme the 
appropriate coefficients are given in Table $21$. Numerically we have 
\begin{eqnarray}
\left. \Sigma^{T}_{(1)}(p,q) \right|_{\alts} &=& -~ 1 ~+~ O(a^3) \nonumber \\
\left. \Sigma^{T}_{(2)}(p,q) \right|_{\alts} &=& [ 1.0417366 \alpha 
+ 3.1252097 ] a \nonumber \\
&& +~ [ 76.3969887 + 19.1449675 \alpha + 5.4041904 \alpha^2 + 0.7813024
\alpha^3 \nonumber \\ 
&& ~~~~-~ [ 5.5559283 + 1.1574851 \alpha ] \Nf ] a^2 ~+~ O(a^3) 
\nonumber \\
\left. \Sigma^{T}_{(3)}(p,q) \right|_{\alts} &=& 
\left. \Sigma^{T}_{(6)}(p,q) \right|_{\alts} \nonumber \\
&=& [ 0.3056953 \alpha - 0.3056953 ] a \nonumber \\
&& -~ [ 1.8386328 - 5.7159421 \alpha - 1.5348987 \alpha^2 - 0.2292715
\alpha^3 \nonumber \\ 
&& ~~~~-~ [ 0.3396615 - 0.3396615 \alpha ] \Nf ] a^2 ~+~ O(a^3) \nonumber \\
\left. \Sigma^{T}_{(4)}(p,q) \right|_{\alts} &=& 
\left. \Sigma^{T}_{(5)}(p,q) \right|_{\alts} \nonumber \\
&=& [ 0.1528477 \alpha - 0.1528477 ] a \nonumber \\
&& -~ [ 0.9193164 - 2.8579710 \alpha - 0.7674493 \alpha^2 - 0.1146357 \alpha^3 
\nonumber \\ 
&& ~~~~-~ [ 0.1698307 - 0.1698307 \alpha ] \Nf ] a^2 ~+~ O(a^3) 
\nonumber \\
\left. \Sigma^{T}_{(7)}(p,q) \right|_{\alts} &=& [ 0.6113907 \alpha 
- 0.6113907 ] a \nonumber \\
&& -~ [ 3.6772656 - 11.4318841 \alpha - 3.0697974 \alpha^2 - 0.4585430
\alpha^3 \nonumber \\ 
&& ~~~~-~ [ 0.6793229 - 0.6793229 \alpha ] \Nf ] a^2 ~+~ O(a^3) 
\nonumber \\
\left. \Sigma^{T}_{(8)}(p,q) \right|_{\alts} &=& -~ [ 1.0417366 \alpha 
+ 3.1252097 ] a \nonumber \\
&& -~ [ 76.3969887 + 19.1449675 \alpha + 5.4041904 \alpha^2 + 0.7813024
\alpha^3 \nonumber \\ 
&& ~~~~-~ [ 5.5559283 + 1.1574851 \alpha ] \Nf ] a^2 ~+~ O(a^3) ~. 
\end{eqnarray}
For the $\Nf$ independent part at two loops the numerical differences between
these amplitudes and those of (\ref{tensampl}) is insignificant. 

\sect{Conversion functions.}

In this section we record the conversion functions for changing from one scheme
to another for the various operators considered here. For an excellent 
background to this see, for example, \cite{38}. The conversion functions, 
$C_i(a,\alpha)$, are defined from the explicit forms of the appropriate 
renormalization constants in each scheme. So, for instance,  
\begin{equation}
C^S(a,\alpha) ~=~ \frac{Z^S_{\mbox{\footnotesize{RI$^\prime$/SMOM}}}}
{Z^S_{\mbox{\footnotesize{$\MSbar$}}}} ~~~,~~~
C^V(a,\alpha) ~=~ \frac{Z^V_{\mbox{\footnotesize{RI$^\prime$/SMOM}}}}
{Z^V_{\mbox{\footnotesize{$\MSbar$}}}} ~~~,~~~
C^T(a,\alpha) ~=~ \frac{Z^T_{\mbox{\footnotesize{RI$^\prime$/SMOM}}}}
{Z^T_{\mbox{\footnotesize{$\MSbar$}}}} ~. 
\label{consvt}
\end{equation}
However, in deriving the explicit expressions in each case one must make a
choice of scheme in which to express all the parameters themselves in. Here, we
will use the $\MSbar$ scheme as the base for the variables. So that in the 
conversion functions, (\ref{consvt}), $\alpha$ and $a$ are $\MSbar$ parameters.
Further, in practical terms this means that the RI${}^\prime$/SMOM scheme 
renormalization constant has been determined as a function of the 
RI${}^\prime$/SMOM scheme version of $\alpha$ and $a$. These have first to be 
mapped to their $\MSbar$ equivalent before the ratios in (\ref{consvt}) can be
computed. Otherwise one might find that the conversion functions are not finite
in four dimensions as they ought to be. With our definition of the 
RI${}^\prime$/SMOM scheme based on the RI${}^\prime$ scheme definition of
$\alpha$, \cite{13}, we have
\begin{eqnarray}
C^S(a,\alpha) &=& 1 ~+~ \left[ ( 3 \alpha + 9) \psi^\prime(\third) 
- ( 6 + 2 \alpha) \pi^2 - 9 \alpha - 36 \right] \frac{C_F a}{9} \nonumber \\
&& +~ \left[ \left[ 
( 72 \alpha^2 + 432 \alpha + 936 ) (\psi^\prime(\third))^2 
- ( 1248 + 576 \alpha + 96 \alpha^2 ) \psi^\prime(\third) \pi^2
\right. \right. \nonumber \\
&& \left. \left. ~~~~~
-~ ( 5616 + 864 \alpha + 432 \alpha^2 ) \psi^\prime(\third) 
- ( 72 + 36 \alpha ) \psi^{\prime\prime\prime}(\third) 
\right. \right. \nonumber \\
&& \left. \left. ~~~~~
-~ 15552 s_2(\pisix) 
+ 31104 s_2(\pitwo) 
+ 25920 s_3(\pisix) 
- 20736 s_3(\pitwo) 
\right. \right. \nonumber \\
&& \left. \left. ~~~~~
+~ ( 32 \alpha^2 + 288 \alpha + 608 ) \pi^4 
+ ( 288 \alpha^2 + 576 \alpha + 3744 ) \pi^2 
\right. \right. \nonumber \\
&& \left. \left. ~~~~~
+~ 648 \alpha^2 + 2592 \alpha + 1539 
+ ( 648 \alpha + 1944 ) \Sigma
\right. \right. \nonumber \\
&& \left. \left. ~~~~~
+~ 2592 \zeta(3)
- 108 \frac{\ln^2(3) \pi}{\sqrt{3}}
+ 1296 \frac{\ln(3) \pi}{\sqrt{3}}
+ 116 \frac{\pi^3}{\sqrt{3}}
\right] C_F \right. \nonumber \\
&& ~~~~~+ \left. \left[ 
192 \psi^\prime(\third) \pi^2 
- 144 (\psi^\prime(\third))^2 
+ ( 162 \alpha^2 + 756 \alpha + 8226 ) \psi^\prime(\third) 
\right. \right. \nonumber \\
&& \left. \left. ~~~~~~~~~
-~ ( 45 + 9 \alpha ) \psi^{\prime\prime\prime}(\third) 
+ 7776 s_2(\pisix) 
- 15552 s_2(\pitwo) 
- 12960 s_3(\pisix) 
\right. \right. \nonumber \\
&& \left. \left. ~~~~~~~~~
+~ 10368 s_3(\pitwo) 
+ ( 24 \alpha + 56 ) \pi^4 
- ( 108 \alpha^2 + 504 \alpha + 5484 ) \pi^2 
\right. \right. \nonumber \\
&& \left. \left. ~~~~~~~~~
-~ 486 \alpha^2 - 2268 \alpha - 34695
+ ( 324 \alpha + 1620 ) \Sigma
+ 6480 \zeta(3)
\right. \right. \nonumber \\
&& \left. \left. ~~~~~~~~~
+~ 54 \frac{\ln^2(3) \pi}{\sqrt{3}}
- 648 \frac{\ln(3) \pi}{\sqrt{3}}
- 58 \frac{\pi^3}{\sqrt{3}}
\right] C_A \right. \nonumber \\
&& ~~~~~+~ \left. \left[ 
960 \pi^2 
- 1440 \psi^\prime(\third) 
+ 8964
\right] T_F \Nf 
\right] \frac{C_F a^2}{648} ~+~ O(a^3) ~.
\end{eqnarray} 
We have checked that the Landau gauge part of this agrees with \cite{17,18} but
the $\alpha$ dependence differs since as we have noted we have renormalized 
$\alpha$ in accordance with \cite{13}. For the vector the situation is 
effectively trivial due to the Slavnov-Taylor identity and so to the order we
have computed 
\begin{equation}
C^V(a,\alpha) ~=~ 1 ~+~ O(a^3) ~.
\end{equation}
For the tensor operator we have 
\begin{eqnarray}
C^T(a,\alpha) &=& 1 ~+~ \left[ ( 3 \alpha - 3 ) \psi^\prime(\third) 
+ ( 2 - 2 \alpha ) \pi^2 - 3 \alpha + 12 \right] \frac{C_F a}{9} \nonumber \\
&& +~ \left[ \left[ 
( 216 \alpha^2 - 432 \alpha + 504 ) (\psi^\prime(\third))^2 
+ ( 576 \alpha - 288 \alpha^2 - 672 ) \psi^\prime(\third) \pi^2
\right. \right. \nonumber \\
&& \left. \left. ~~~~~
+~ ( 13824 \alpha - 432 \alpha^2 - 40176 ) \psi^\prime(\third) 
+ ( 144 - 108 \alpha ) \psi^{\prime\prime\prime}(\third) 
\right. \right. \nonumber \\
&& \left. \left. ~~~~~
+~ ( 62208 \alpha - 233280 ) s_2(\pisix) 
+ ( 466560 - 124416 \alpha ) s_2(\pitwo) 
\right. \right. \nonumber \\
&& \left. \left. ~~~~~
+~ ( 388800 - 103680 \alpha ) s_3(\pisix) 
+ ( 82944 \alpha - 311040 ) s_3(\pitwo) 
\right. \right. \nonumber \\
&& \left. \left. ~~~~~
+~ ( 96 \alpha^2 + 96 \alpha - 160 ) \pi^4 
+ ( 288 \alpha^2 - 9216 \alpha + 26784 ) \pi^2 
\right. \right. \nonumber \\
&& \left. \left. ~~~~~
+~ 216 \alpha^2 - 1728 \alpha - 45063
+ ( 1944 \alpha - 1944 ) \Sigma
\right. \right. \nonumber \\
&& \left. \left. ~~~~~
+~ ( 5184 \alpha + 23328 ) \zeta(3)
+ ( 432 \alpha - 1620 ) \frac{\ln^2(3) \pi}{\sqrt{3}}
\right. \right. \nonumber \\
&& \left. \left. ~~~~~
+~ ( 19440 - 5184 \alpha ) \frac{\ln(3) \pi}{\sqrt{3}}
+ ( 1740 - 464 \alpha ) \frac{\pi^3}{\sqrt{3}}
\right] C_F \right. \nonumber \\
&& ~~~~~+ \left. \left[ 
192 \psi^\prime(\third) \pi^2 
- 144 (\psi^\prime(\third))^2 
+ ( 486 \alpha^2 - 1620 \alpha + 13590 ) \psi^\prime(\third) 
\right. \right. \nonumber \\
&& \left. \left. ~~~~~~~~~
+~ ( 81 - 27 \alpha ) \psi^{\prime\prime\prime}(\third) 
+ ( 124416 - 23328 \alpha ) s_2(\pisix) 
\right. \right. \nonumber \\
&& \left. \left. ~~~~~~~~~
+~ ( 46656 \alpha - 248832 ) s_2(\pitwo) 
+ ( 38880 \alpha - 207360 ) s_3(\pisix) 
\right. \right. \nonumber \\
&& \left. \left. ~~~~~~~~~
+~ ( 165888 - 31104 \alpha ) s_3(\pitwo) 
+ ( 72 \alpha - 280 ) \pi^4 
\right. \right. \nonumber \\
&& \left. \left. ~~~~~~~~~
-~ ( 324 \alpha^2 - 1080 \alpha + 9060 ) \pi^2 
- 486 \alpha^2 - 2268 \alpha + 76419
\right. \right. \nonumber \\
&& \left. \left. ~~~~~~~~~
+~ ( 972 \alpha - 1620 ) \Sigma
- 32400 \zeta(3)
+ ( 864 - 162 \alpha ) \frac{\ln^2(3) \pi}{\sqrt{3}}
\right. \right. \nonumber \\
&& \left. \left. ~~~~~~~~~
+~ ( 1944 \alpha - 10368 ) \frac{\ln(3) \pi}{\sqrt{3}}
+ ( 174 \alpha - 928 ) \frac{\pi^3}{\sqrt{3}}
\right] C_A \right. \nonumber \\
&& ~~~~~+~ \left. \left[ 
1440 \psi^\prime(\third) 
- 960 \pi^2 
- 17028
\right] T_F \Nf 
\right] \frac{C_F a^2}{1944} ~+~ O(a^3)
\end{eqnarray} 
where $C^T(a,0)$ agrees with \cite{17,18}. Numerically we have 
\begin{eqnarray}
C^S(a,\alpha) &=& 1 ~+~ [ 0.2292715 \alpha - 0.6455188 ] a \nonumber \\
&& +~ [ 0.5684263 \alpha^2 + 4.5546643 \alpha - 22.6076874 + 4.0135395 \Nf ]
a^2 ~+~ O(a^3) \nonumber \\
C^V(a,\alpha) &=& 1 ~+~ O(a^3) \nonumber \\
C^T(a,\alpha) &=& 1 ~+~ [ 1.1181604 \alpha + 0.2151729 ] a \nonumber \\
&& +~ [ 3.7661435 \alpha^2 + 10.8071579 \alpha + 43.4302495 - 4.1032786 \Nf ]
a^2 \nonumber \\ 
&& +~ O(a^3) ~.
\end{eqnarray} 
For the tensor current in order to compare with the alternative scheme we have
the alternative scheme conversion function to $\MSbar$  
\begin{eqnarray}
\left. C^T(a,\alpha) \right|_{\alts} &=& 1 ~+~ \left[ ( 15 \alpha - 15 ) 
\psi^\prime(\third) + ( 10 - 10 \alpha ) \pi^2 - 27 \alpha + 54 \right] 
\frac{C_F a}{27} \nonumber \\
&& +~ \left[ \left[ 
( 1800 \alpha^2 - 3600 \alpha + 4392 ) (\psi^\prime(\third))^2 
+ ( 4800 \alpha - 2400 \alpha^2 - 5856 ) \psi^\prime(\third) \pi^2
\right. \right. \nonumber \\
&& \left. \left. ~~~~~
+~ ( 58320 \alpha - 6480 \alpha^2 - 137376 ) \psi^\prime(\third) 
+ ( 1296 - 540 \alpha ) \psi^{\prime\prime\prime}(\third) 
\right. \right. \nonumber \\
&& \left. \left. ~~~~~
+~ ( 186624 \alpha - 513216 ) s_2(\pisix) 
+ ( 1026432 - 373248 \alpha ) s_2(\pitwo) 
\right. \right. \nonumber \\
&& \left. \left. ~~~~~
+~ ( 855360 - 311040 \alpha ) s_3(\pisix) 
+ ( 248832 \alpha - 684288 ) s_3(\pitwo) 
\right. \right. \nonumber \\
&& \left. \left. ~~~~~
+~ ( 800 \alpha^2 - 160 \alpha - 1504 ) \pi^4 
+ ( 4320 \alpha^2 - 38880 \alpha + 91584 ) \pi^2 
\right. \right. \nonumber \\
&& \left. \left. ~~~~~
+~ 5832 \alpha^2 - 11664 \alpha - 133893
+ ( 9720 \alpha - 9720 ) \Sigma
\right. \right. \nonumber \\
&& \left. \left. ~~~~~
+~ ( 15552 \alpha + 38880 ) \zeta(3)
+ ( 1296 \alpha - 3564 ) \frac{\ln^2(3) \pi}{\sqrt{3}}
\right. \right. \nonumber \\
&& \left. \left. ~~~~~
+~ ( 42768 - 15552 \alpha ) \frac{\ln(3) \pi}{\sqrt{3}}
+ ( 3828 - 1392 \alpha ) \frac{\pi^3}{\sqrt{3}}
\right] C_F \right. \nonumber \\
&& ~~~~~+ \left. \left[ 
1728 \psi^\prime(\third) \pi^2 
- 1296 (\psi^\prime(\third))^2 
+ ( 2430 \alpha^2 - 324 \alpha + 16542 ) \psi^\prime(\third) 
\right. \right. \nonumber \\
&& \left. \left. ~~~~~~~~~
-~ ( 135 + 135 \alpha ) \psi^{\prime\prime\prime}(\third) 
+ ( 139968 - 69984 \alpha ) s_2(\pisix) 
\right. \right. \nonumber \\
&& \left. \left. ~~~~~~~~~
+~ ( 139968 \alpha - 279936 ) s_2(\pitwo) 
+ ( 116640 \alpha - 233280 ) s_3(\pisix) 
\right. \right. \nonumber \\
&& \left. \left. ~~~~~~~~~
+~ ( 186624 - 93312 \alpha ) s_3(\pitwo) 
+ ( 360 \alpha - 216 ) \pi^4 
\right. \right. \nonumber \\
&& \left. \left. ~~~~~~~~~
-~ ( 1620 \alpha^2 - 216 \alpha + 11028 ) \pi^2 
- 4374 \alpha^2 - 20412 \alpha + 251397
\right. \right. \nonumber \\
&& \left. \left. ~~~~~~~~~
+~ ( 4860 \alpha - 6804 ) \Sigma
- 69984 \zeta(3)
+ ( 972 - 486 \alpha ) \frac{\ln^2(3) \pi}{\sqrt{3}}
\right. \right. \nonumber \\
&& \left. \left. ~~~~~~~~~
+~ ( 5832 \alpha - 11664 ) \frac{\ln(3) \pi}{\sqrt{3}}
+ ( 522 \alpha - 1044 ) \frac{\pi^3}{\sqrt{3}}
\right] C_A \right. \nonumber \\
&& ~~~~~+~ \left. \left[ 
7200 \psi^\prime(\third) 
- 4800 \pi^2 
- 59724
\right] T_F \Nf 
\right] \frac{C_F a^2}{5832} ~+~ O(a^3) ~.
\end{eqnarray} 
Numerically this equates to
\begin{eqnarray}
\left. C^T(a,\alpha) \right|_{\alts} &=& 1 ~+~ [ 1.2710080 \alpha 
+ 0.0623253 ] a \nonumber \\
&& +~ [ 4.4752296 \alpha^2 + 12.2915908 \alpha + 42.4780444 - 3.9334478 \Nf ]
a^2 \nonumber \\ 
&& +~ O(a^3) ~.
\end{eqnarray} 
In \cite{19} it was noted that the Landau gauge one loop correction to the 
conversion function in this alternative scheme was significantly smaller than 
that of the scheme of \cite{17,18}. However, with the two loop computations 
complete it transpires that the two loop terms are comparable.  

One of the roles of the conversion function is to allow one to map anomalous
dimensions from one renormalization scheme to another. For instance, in the
tensor case the two loop RI${}^\prime$/SMOM scheme anomalous dimension, 
(\ref{tens2}), can be deduced from the $\MSbar$ version from $C^T(a,\alpha)$ 
using 
\begin{eqnarray}
\gamma^T_{\mbox{\footnotesize{RI$^\prime$/SMOM}}}
\left(a_{\mbox{\footnotesize{RI$^\prime$}}},
\alpha_{\mbox{\footnotesize{RI$^\prime$}}}\right) &=&
\gamma^T_{\mbox{\footnotesize{$\MSbar$}}}
\left(a_{\mbox{\footnotesize{$\MSbar$}}}\right) ~-~
\beta\left(a_{\mbox{\footnotesize{$\MSbar$}}}\right)
\frac{\partial ~}{\partial a_{\mbox{\footnotesize{$\MSbar$}}}}
\ln C^T\left(a_{\mbox{\footnotesize{$\MSbar$}}},
\alpha_{\mbox{\footnotesize{$\MSbar$}}}\right) \nonumber \\
&& -~ \alpha_{\mbox{\footnotesize{$\MSbar$}}}
\gamma^{\mbox{\footnotesize{$\MSbar$}}}_\alpha
\left(a_{\mbox{\footnotesize{$\MSbar$}}}\right)
\frac{\partial ~}{\partial \alpha_{\mbox{\footnotesize{$\MSbar$}}}}
\ln C^T\left(a_{\mbox{\footnotesize{$\MSbar$}}},
\alpha_{\mbox{\footnotesize{$\MSbar$}}}\right)
\end{eqnarray}
where in this case we have been careful in making it explicit what scheme the
variables are in. We have checked that the RI${}^\prime$/SMOM scheme anomalous 
dimension of both the scalar and tensor follow precisely from the respective
conversion functions. However, given the way the coupling constants appear 
throughout it is only the one loop part of the conversion function which is 
used in this. Therefore, given the fact that the three loop $\MSbar$ scalar and
tensor operator anomalous dimensions are known then it is possible to deduce 
the three loop RI${}^\prime$/SMOM for the tensor current. We have verified that
the three loop Landau gauge results of \cite{17} are obtained.
 
\sect{Discussion.}

We have computed the full two loop form of the Green's function of Figure $1$
for three quark currents in both the $\MSbar$ and RI${}^\prime$/SMOM
renormalization schemes. The amplitudes for the former case will assist with
matching lattice computations of the same Green's functions at high energy once
the numerical results are expressed in the same scheme. The latter amplitudes
will play a similar role but in the case where the renormalization scheme used
on the lattice is the same as the RI${}^\prime$/SMOM scheme defined in
\cite{16,17,18}. However, as we have discussed, in the case of the tensor 
current, there are in principle different ways of defining an
RI${}^\prime$/SMOM type scheme. This is because with free Lorentz indices in 
the operator there is more than one amplitude with respect to a basis of 
Lorentz tensors. The tensor bases we have introduced here for the various 
operators are by no means unique. Since the renormalization constant is defined
in relation to a basis tensor or tensors, then a change of basis would lead to
different numerical structure in the renormalization constants themselves as
well as the conversion functions. Whilst this degree of ambiguity in defining
an RI${}^\prime$/SMOM scheme for these currents may appear to be an issue, it
may in fact be possible to exploit it to render corrections in, say, the 
conversion functions such that they are significantly small. Therefore, one in
principle can improve the accuracy of any measurements. For the tensor case it
was noted in \cite{19} that an alternative scheme in the tensor current case
could produce a smaller one loop correction than that of the original work of 
\cite{16,17,18}. However, our two loop computation demonstrated that this is
not retained at that order. Though one could conceive of a way of producing a 
smaller correction with the absorption of an appropriate finite part from 
another amplitude by a basis redefinition. Whilst we have focused on quark 
currents the programmes we have developed can be used to consider the same 
problem for the low moment operators used in deep inelastic scattering, 
\cite{20}. The one loop analysis was given in \cite{19} for the operator sets 
labelled $W_2$ and $W_3$. However, the treatment of the vector current here 
underpins any two loop extension of \cite{19}. This is because within each of 
the sets $W_2$ and $W_3$ the vector current is present. Though as the deep 
inelastic scattering operators involve the covariant derivative the presence of
the vector current is through its divergence since there is operator mixing. As
a result of this the renormalization of the resident vector current still has 
to be consistent with the Slavnov-Taylor identity in the RI${}^\prime$/SMOM 
renormalization scheme. Therefore, we have laid the groundwork for that in this
article by treating the vector current and its amplitudes when inserted in a 
quark $2$-point function at length.

\vspace{1cm}
\noindent
{\bf Acknowledgement.} The author thanks Dr. P.E.L. Rakow and Prof. C.T.C. 
Sachrajda for useful discussions.


\vspace{3cm}
{\begin{table}[ht]
\begin{center}
\begin{tabular}{|c||r|r||r|r|}
\hline
$a^{(1)}_n$ & $\left. c^{S,(1)}_{(1)\,n} \right|_{\MSbars}$ & 
$\left. c^{S,(1)}_{(2)\,n} \right|_{\MSbars}$ & 
$c^{S,(1)}_{(1)\,n}$ & $c^{S,(1)}_{(2)\,n}$ \\
\hline
$1$ & $- 4$ & $0$ & $0$ & $0$ \\
$\pi^2 \alpha$ & $- 2/9$ & $- 4/27$ & $0$ & $- 4/27$ \\
$\alpha$ & $- 2$ & $0$ & $0$ & $0$ \\
$\pi^2$ & $0$ & $4/27$ & $0$ & $4/27$ \\
$\psi^\prime(1/3)$ & $1$ & $- 2/9$ & $0$ & $- 2/9$ \\
$\psi^\prime(1/3) \alpha$ & $1/3$ & $2/9$ & $0$ & $2/9$ \\
\hline
\end{tabular}
\end{center}
\begin{center}
{Table $1$. $\MSbar$ and RI${}^\prime$/SMOM scheme coefficients of $C_F$ for 
one loop $S$ amplitudes.}
\end{center}
\end{table}}

\vspace{4cm}
{\begin{table}[hb]
\begin{center}
\begin{tabular}{|c||r|r|r|r|r|r|}
\hline
$a^{(21)}_n$ & $\left. c^{S,(21)}_{(1)\,n} \right|_{\MSbars}$ & 
$\left. c^{S,(21)}_{(2)\,n} \right|_{\MSbars}$ & $c^{S,(21)}_{(2)\,n}$ \\
\hline
$1$ & $52/3$ & $0$ & $0$ \\
$\pi^2 \alpha$ & $0$ & $0$ & $80/243$ \\
$\alpha$ & $0$ & $0$ & $0$ \\
$\pi^2$ & $40/27$ & $- 32/243$ & $- 32/243$ \\
$\psi^\prime(1/3)$ & $- 20/9$ & $16/81$ & $16/81$ \\
$\psi^\prime(1/3) \alpha$ & $0$ & $0$ & $- 40/81$ \\
\hline
\end{tabular}
\end{center}
\begin{center}
{Table $2$. $\MSbar$ and RI${}^\prime$/SMOM coefficients of $C_F T_F \Nf$ for
two loop $S$ amplitudes.}
\end{center}
\end{table}}

\clearpage

{\begin{table}[ht]
\begin{center}
\begin{tabular}{|c||r|r|r|}
\hline
$a^{(22)}_n$ & $\left. c^{S,(22)}_{(1)\,n} \right|_{\MSbars}$ & 
$\left. c^{S,(22)}_{(2)\,n} \right|_{\MSbars}$ & $c^{S,(22)}_{(2)\,n}$ \\
\hline
$1$ & $- 1531/24$ & $0$ & $0$ \\
$\pi^2 \alpha$ & $- 7/9$ & $28/27$ & $155/243$ \\
$\pi^4 \alpha$ & $1/27$ & $- 2/81$ & $- 2/81$ \\
$\zeta(3) \alpha$ & $3$ & $- 2/3$ & $- 2/3$ \\
$\Sigma \alpha$ & $1/2$ & $1/3$ & $1/3$ \\
$\alpha$ & $- 10$ & $0$ & $0$ \\
$\pi^2 \alpha^2$ & $- 1/6$ & $- 1/27$ & $- 1/9$ \\
$\alpha^2$ & $- 15/8$ & $0$ & $0$ \\
$\pi^2 \alpha^3$ & $0$ & $0$ & $- 1/27$ \\
$\alpha^3$ & $0$ & $0$ & $0$ \\
$\pi^2$ & $- 457/54$ & $- 1691/243$ & $- 1691/243$ \\
$\pi^4$ & $7/81$ & $- 26/81$ & $- 26/81$ \\
$\zeta(3)$ & $13$ & $- 38/3$ & $- 38/3$ \\
$\Sigma$ & $5/2$ & $- 1$ & $- 1$ \\
$s_2(\pi/6)$ & $12$ & $40$ & $40$ \\
$s_2(\pi/6) \alpha$ & $0$ & $- 8$ & $- 8$ \\
$s_2(\pi/2)$ & $- 24$ & $- 80$ & $- 80$ \\
$s_2(\pi/2) \alpha$ & $0$ & $16$ & $16$ \\
$s_3(\pi/6)$ & $- 20$ & $- 200/3$ & $- 200/3$ \\
$s_3(\pi/6) \alpha$ & $0$ & $40/3$ & $40/3$ \\
$s_3(\pi/2)$ & $16$ & $160/3$ & $160/3$ \\
$s_3(\pi/2) \alpha$ & $0$ & $- 32/3$ & $- 32/3$ \\
$\psi^\prime(1/3)$ & $457/36$ & $1691/162$ & $1691/162$ \\
$\psi^\prime(1/3) \alpha$ & $7/6$ & $- 14/9$ & $- 155/162$ \\
$\psi^\prime(1/3) \alpha^2$ & $1/4$ & $1/18$ & $1/6$ \\
$\psi^\prime(1/3) \alpha^3$ & $0$ & $0$ & $1/18$ \\
$\psi^\prime(1/3) \pi^2 $ & $8/27$ & $16/27$ & $16/27$ \\
$(\psi^\prime(1/3))^2$ & $- 2/9$ & $- 4/9$ & $- 4/9$ \\
$\psi^{\prime\prime\prime}(1/3)$ & $- 5/72$ & $5/108$ & $5/108$ \\
$\psi^{\prime\prime\prime}(1/3) \alpha$ & $- 1/72$ & $1/108$ & $1/108$ \\
$\pi^3 \alpha/\sqrt{3}$ & $0$ & $29/486$ & $29/486$ \\
$\pi^3/\sqrt{3}$ & $- 29/324$ & $- 145/486$ & $- 145/486$ \\
$\pi \ln(3) \alpha/\sqrt{3}$ & $0$ & $2/3$ & $2/3$ \\
$\pi \ln(3)/\sqrt{3}$ & $- 1$ & $- 10/3$ & $- 10/3$ \\
$\pi (\ln(3))^2 \alpha/\sqrt{3}$ & $0$ & $- 1/18$ & $- 1/18$ \\
$\pi (\ln(3))^2/\sqrt{3}$ & $1/12$ & $5/18$ & $5/18$ \\
\hline
\end{tabular}
\end{center}
\begin{center}
{Table $3$. $\MSbar$ and RI${}^\prime$/SMOM coefficients of $C_F C_A$ for two 
loop $S$ amplitudes.}
\end{center}
\end{table}}

\clearpage

{\begin{table}[ht]
\begin{center}
\begin{tabular}{|c||r|r|r|}
\hline
$a^{(23)}_n$ & $\left. c^{S,(23)}_{(1)\,n} \right|_{\MSbars}$ & 
$\left. c^{S,(23)}_{(2)\,n} \right|_{\MSbars}$ & $c^{S,(23)}_{(2)\,n}$ \\
\hline
$1$ & $- 13$ & $0$ & $0$ \\
$\pi^2 \alpha$ & $- 26/9$ & $- 92/27$ & $- 28/9$ \\
$\pi^4 \alpha$ & $4/27$ & $8/81$ & $40/243$ \\
$\zeta(3) \alpha$ & $0$ & $8/3$ & $8/3$ \\
$\Sigma \alpha$ & $1$ & $2/3$ & $2/3$ \\
$\alpha$ & $- 8$ & $0$ & $0$ \\
$\pi^2 \alpha^2$ & $- 2/9$ & $- 8/27$ & $0$ \\
$\pi^4 \alpha^2$ & $0$ & $0$ & $8/243$ \\
$\alpha^2$ & $- 1$ & $0$ & $0$ \\
$\pi^2$ & $4/9$ & $580/27$ & $188/9$ \\
$\pi^4$ & $40/81$ & $64/81$ & $56/81$ \\
$\zeta(3)$ & $4$ & $24$ & $24$ \\
$\Sigma$ & $3$ & $- 2/3$ & $- 2/3$ \\
$s_2(\pi/6)$ & $- 24$ & $- 192$ & $- 192$ \\
$s_2(\pi/6) \alpha$ & $0$ & $32$ & $32$ \\
$s_2(\pi/2)$ & $48$ & $384$ & $384$ \\
$s_2(\pi/2) \alpha$ & $0$ & $- 64$ & $- 64$ \\
$s_3(\pi/6)$ & $40$ & $320$ & $320$ \\
$s_3(\pi/6) \alpha$ & $0$ & $- 160/3$ & $- 160/3$ \\
$s_3(\pi/2)$ & $- 32$ & $- 256$ & $- 256$ \\
$s_3(\pi/2) \alpha$ & $0$ & $128/3$ & $128/3$ \\
$\psi^\prime(1/3)$ & $- 2/3$ & $- 290/9$ & $- 94/3$ \\
$\psi^\prime(1/3) \pi^2 \alpha$ & $0$ & $0$ & $- 16/81$ \\
$\psi^\prime(1/3) \alpha$ & $13/3$ & $46/9$ & $14/3$ \\
$\psi^\prime(1/3) \pi^2 \alpha^2 $ & $0$ & $0$ & $- 8/81$ \\
$\psi^\prime(1/3) \alpha^2 $ & $1/3$ & $4/9$ & $0$ \\
$\psi^\prime(1/3) \pi^2 $ & $- 16/27$ & $- 32/27$ & $- 8/9$ \\
$(\psi^\prime(1/3))^2$ & $4/9$ & $8/9$ & $2/3$ \\
$(\psi^\prime(1/3))^2 \alpha$ & $0$ & $0$ & $4/27$ \\
$(\psi^\prime(1/3))^2 \alpha^2$ & $0$ & $0$ & $2/27$ \\
$\psi^{\prime\prime\prime}(1/3)$ & $- 1/9$ & $- 4/27$ & $- 4/27$ \\
$\psi^{\prime\prime\prime}(1/3) \alpha$ & $- 1/18$ & $- 1/27$ & $- 1/27$ \\
$\pi^3 \alpha/\sqrt{3}$ & $0$ & $- 58/243$ & $- 58/243$ \\
$\pi^3/\sqrt{3}$ & $29/162$ & $116/81$ & $116/81$ \\
$\pi \ln(3) \alpha/\sqrt{3}$ & $0$ & $- 8/3$ & $- 8/3$ \\
$\pi \ln(3)/\sqrt{3}$ & $2$ & $16$ & $16$ \\
$\pi (\ln(3))^2 \alpha/\sqrt{3}$ & $0$ & $2/9$ & $2/9$ \\
$\pi (\ln(3))^2/\sqrt{3}$ & $- 1/6$ & $- 4/3$ & $- 4/3$ \\
\hline
\end{tabular}
\end{center}
\begin{center}
{Table $4$. $\MSbar$ and RI${}^\prime$/SMOM coefficients of $C_F^2$ for two 
loop $S$ amplitudes.}
\end{center}
\end{table}}

\clearpage

{\begin{table}[ht]
\begin{center}
\begin{tabular}{|c||r|r|r|r|}
\hline
$a^{(1)}_n$ & $\left. c^{V,(1)}_{(1)\,n} \right|_{\MSbars}$ & 
$\left. c^{V,(1)}_{(2)\,n} \right|_{\MSbars}$ & 
$\left. c^{V,(1)}_{(3)\,n} \right|_{\MSbars}$ & 
$\left. c^{V,(1)}_{(6)\,n} \right|_{\MSbars}$ \\
\hline
$1$ & $2$ & $8/3$ & $4/3$ & $0$ \\
$\pi^2 \alpha$ & $- 8/27$ & $- 8/27$ & $- 8/27$ & $0$ \\
$\alpha$ & $- 2$ & $- 4/3$ & $- 2/3$ & $0$ \\
$\pi^2$ & $4/27$ & $8/27$ & $0$ & $8/27$ \\
$\psi^\prime(1/3)$ & $- 2/9$ & $- 4/9$ & $0$ & $- 4/9$ \\
$\psi^\prime(1/3) \alpha$ & $4/9$ & $4/9$ & $4/9$ & $0$ \\
\hline
\end{tabular}
\end{center}
\begin{center}
{Table $5$. $\MSbar$ coefficients of $C_F$ for one loop $V$ amplitudes.}
\end{center}
\end{table}}

\vspace{4cm}
{\begin{table}[hb]
\begin{center}
\begin{tabular}{|c||r|r|r|r|}
\hline
$a^{(21)}_n$ & $\left. c^{V,(21)}_{(1)\,n} \right|_{\MSbars}$ & 
$\left. c^{V,(21)}_{(2)\,n} \right|_{\MSbars}$ & 
$\left. c^{V,(21)}_{(3)\,n} \right|_{\MSbars}$ & 
$\left. c^{V,(21)}_{(6)\,n} \right|_{\MSbars}$ \\
\hline
$1$ & $- 53/18$ & $- 232/27$ & $- 116/27$ & $0$ \\
$\pi^2 \alpha$ & $0$ & $0$ & $0$ & $0$ \\
$\alpha$ & $0$ & $0$ & $0$ & $0$ \\
$\pi^2$ & $- 152/243$ & $- 304/243$ & $0$ & $- 208/243$ \\
$\psi^\prime(1/3)$ & $76/81$ & $152/81$ & $0$ & $104/81$ \\
$\psi^\prime(1/3) \alpha$ & $0$ & $0$ & $0$ & $0$ \\
\hline
\end{tabular}
\end{center}
\begin{center}
{Table $6$. $\MSbar$ coefficients of $C_F T_F \Nf$ for two loop $V$ 
amplitudes.}
\end{center}
\end{table}}

\clearpage

{\begin{table}[ht]
\begin{center}
\begin{tabular}{|c||r|r|r|r|}
\hline
$a^{(22)}_n$ & $\left. c^{V,(22)}_{(1)\,n} \right|_{\MSbars}$ & 
$\left. c^{V,(22)}_{(2)\,n} \right|_{\MSbars}$ & 
$\left. c^{V,(22)}_{(3)\,n} \right|_{\MSbars}$ & 
$\left. c^{V,(22)}_{(6)\,n} \right|_{\MSbars}$ \\
\hline
$1$ & $169/18$ & $707/54$ & $707/54$ & $0$ \\
$\pi^2 \alpha$ & $- 10/27$ & $- 28/27$ & $8/27$ & $2/9$ \\
$\pi^4 \alpha$ & $4/81$ & $4/81$ & $4/81$ & $- 4/81$ \\
$\zeta(3) \alpha$ & $3$ & $0$ & $0$ & $- 2/3$ \\
$\Sigma \alpha$ & $2/3$ & $2/3$ & $2/3$ & $0$ \\
$\alpha$ & $- 10$ & $- 14/3$ & $- 7/3$ & $0$ \\
$\pi^2 \alpha^2$ & $- 2/9$ & $- 2/9$ & $- 2/9$ & $2/27$ \\
$\alpha^2$ & $- 15/8$ & $- 1$ & $- 1/2$ & $0$ \\
$\pi^2 \alpha^3$ & $0$ & $0$ & $0$ & $0$ \\
$\alpha^3$ & $0$ & $0$ & $0$ & $0$ \\
$\pi^2$ & $- 236/243$ & $266/243$ & $- 82/27$ & $- 1672/243$ \\
$\pi^4$ & $- 10/81$ & $- 32/243$ & $- 28/243$ & $- 32/81$ \\
$\zeta(3)$ & $- 2/3$ & $- 2$ & $- 16/3$ & $- 28/3$ \\
$\Sigma$ & $- 2/3$ & $- 4/3$ & $0$ & $- 4/3$ \\
$s_2(\pi/6)$ & $10$ & $0$ & $20$ & $68$ \\
$s_2(\pi/6) \alpha$ & $- 6$ & $0$ & $- 12$ & $4$ \\
$s_2(\pi/2)$ & $- 20$ & $0$ & $- 40$ & $- 136$ \\
$s_2(\pi/2) \alpha$ & $12$ & $0$ & $24$ & $- 8$ \\
$s_3(\pi/6)$ & $- 50/3$ & $0$ & $- 100/3$ & $- 340/3$ \\
$s_3(\pi/6) \alpha$ & $10$ & $0$ & $20$ & $- 20/3$ \\
$s_3(\pi/2)$ & $40/3$ & $0$ & $80/3$ & $272/3$ \\
$s_3(\pi/2) \alpha$ & $- 8$ & $0$ & $- 16$ & $16/3$ \\
$\psi^\prime(1/3)$ & $118/81$ & $- 133/81$ & $41/9$ & $836/81$ \\
$\psi^\prime(1/3) \alpha$ & $5/9$ & $14/9$ & $- 4/9$ & $- 1/3$ \\
$\psi^\prime(1/3) \alpha^2$ & $1/3$ & $1/3$ & $1/3$ & $- 1/9$ \\
$\psi^\prime(1/3) \alpha^3$ & $0$ & $0$ & $0$ & $0$ \\
$\psi^\prime(1/3) \pi^2 $ & $8/27$ & $32/81$ & $16/81$ & $16/27$ \\
$(\psi^\prime(1/3))^2$ & $- 2/9$ & $- 8/27$ & $- 4/27$ & $- 4/9$ \\
$\psi^{\prime\prime\prime}(1/3)$ & $1/108$ & $- 8/27$ & $1/54$ & $2/27$ \\
$\psi^{\prime\prime\prime}(1/3) \alpha$ & $- 1/54$ & $- 1/54$ & $- 1/54$ & $1/54$ \\
$\pi^3 \alpha/\sqrt{3}$ & $29/648$ & $0$ & $29/324$ & $- 29/972$ \\
$\pi^3/\sqrt{3}$ & $- 145/1944$ & $0$ & $- 145/972$ & $- 493/972$ \\
$\pi \ln(3) \alpha/\sqrt{3}$ & $1/2$ & $0$ & $1$ & $- 1/3$ \\
$\pi \ln(3)/\sqrt{3}$ & $- 5/6$ & $0$ & $- 5/3$ & $- 17/3$ \\
$\pi (\ln(3))^2 \alpha/\sqrt{3}$ & $- 1/24$ & $0$ & $- 1/12$ & $1/36$ \\
$\pi (\ln(3))^2/\sqrt{3}$ & $5/72$ & $0$ & $5/36$ & $17/36$ \\
\hline
\end{tabular}
\end{center}
\begin{center}
{Table $7$. $\MSbar$ coefficients of $C_F C_A$ for two loop $V$ amplitudes.}
\end{center}
\end{table}}

\clearpage

{\begin{table}[t]
\begin{center}
\begin{tabular}{|c||r|r|r|r|}
\hline
$a^{(23)}_n$ & $\left. c^{V,(23)}_{(1)\,n} \right|_{\MSbars}$ & 
$\left. c^{V,(23)}_{(2)\,n} \right|_{\MSbars}$ & 
$\left. c^{V,(23)}_{(3)\,n} \right|_{\MSbars}$ & 
$\left. c^{V,(23)}_{(6)\,n} \right|_{\MSbars}$ \\
\hline
$1$ & $- 23/8$ & $- 14/3$ & $- 7/3$ & $0$ \\
$\pi^2 \alpha$ & $- 64/27$ & $- 20/27$ & $- 4$ & $- 4/27$ \\
$\pi^4 \alpha$ & $16/81$ & $16/81$ & $16/81$ & $0$ \\
$\zeta(3) \alpha$ & $4/3$ & $0$ & $8/3$ & $0$ \\
$\Sigma \alpha$ & $4/3$ & $4/3$ & $4/3$ & $0$ \\
$\alpha$ & $4$ & $16/3$ & $8/3$ & $0$ \\
$\pi^2 \alpha^2$ & $- 8/27$ & $- 8/27$ & $- 8/27$ & $0$ \\
$\pi^4 \alpha^2$ & $0$ & $0$ & $0$ & $0$ \\
$\alpha^2$ & $- 1$ & $- 4/3$ & $- 2/3$ & $0$ \\
$\pi^2$ & $122/27$ & $88/27$ & $52/9$ & $179/9$ \\
$\pi^4$ & $- 20/81$ & $- 128/243$ & $8/243$ & $8/27$ \\
$\zeta(3)$ & $- 28/3$ & $- 16$ & $- 8/3$ & $8$ \\
$\Sigma$ & $- 2/3$ & $- 4/3$ & $0$ & $- 4/3$ \\
$s_2(\pi/6)$ & $8$ & $48$ & $- 32$ & $- 144$ \\
$s_2(\pi/6) \alpha$ & $16$ & $0$ & $32$ & $0$ \\
$s_2(\pi/2)$ & $- 16$ & $- 96$ & $64$ & $288$ \\
$s_2(\pi/2) \alpha$ & $- 32$ & $0$ & $- 64$ & $0$ \\
$s_3(\pi/6)$ & $- 40/3$ & $- 80$ & $160/3$ & $240$ \\
$s_3(\pi/6) \alpha$ & $- 80/3$ & $0$ & $- 160/3$ & $0$ \\
$s_3(\pi/2)$ & $32/3$ & $64$ & $- 128/3$ & $- 192$ \\
$s_3(\pi/2) \alpha$ & $64/3$ & $0$ & $128/3$ & $0$ \\
$\psi^\prime(1/3)$ & $- 61/9$ & $- 44/9$ & $- 26/3$ & $- 86/3$ \\
$\psi^\prime(1/3) \pi^2 \alpha$ & $0$ & $0$ & $0$ & $0$ \\
$\psi^\prime(1/3) \alpha$ & $32/9$ & $10/9$ & $6$ & $0$ \\
$\psi^\prime(1/3) \pi^2 \alpha^2 $ & $0$ & $0$ & $0$ & $0$ \\
$\psi^\prime(1/3) \alpha^2 $ & $4/9$ & $4/9$ & $4/9$ & $2/9$ \\
$\psi^\prime(1/3) \pi^2 $ & $- 16/27$ & $- 64/81$ & $- 32/81$ & $- 32/27$ \\
$(\psi^\prime(1/3))^2$ & $4/9$ & $16/27$ & $8/27$ & $8/9$ \\
$(\psi^\prime(1/3))^2 \alpha$ & $0$ & $0$ & $0$ & $0$ \\
$(\psi^\prime(1/3))^2 \alpha^2$ & $0$ & $0$ & $0$ & $0$ \\
$\psi^{\prime\prime\prime}(1/3)$ & $1/6$ & $8/27$ & $1/27$ & $1/27$ \\
$\psi^{\prime\prime\prime}(1/3) \alpha$ & $- 2/27$ & $- 2/27$ & $- 2/27$ & $0$ \\
$\pi^3 \alpha/\sqrt{3}$ & $- 29/243$ & $0$ & $- 58/243$ & $0$ \\
$\pi^3/\sqrt{3}$ & $- 29/486$ & $- 29/81$ & $58/243$ & $29/27$ \\
$\pi \ln(3) \alpha/\sqrt{3}$ & $- 4/3$ & $0$ & $- 8/3$ & $0$ \\
$\pi \ln(3)/\sqrt{3}$ & $- 2/3$ & $- 4$ & $8/3$ & $12$ \\
$\pi (\ln(3))^2 \alpha/\sqrt{3}$ & $1/9$ & $0$ & $2/9$ & $0$ \\
$\pi (\ln(3))^2/\sqrt{3}$ & $1/18$ & $1/3$ & $- 2/9$ & $- 1$ \\
\hline
\end{tabular}
\end{center}
\begin{center}
{Table $8$. $\MSbar$ coefficients of $C_F^2$ for two loop $V$ amplitudes.}
\end{center}
\end{table}}

\clearpage

{\begin{table}[ht]
\begin{center}
\begin{tabular}{|c||r|r|r|r|}
\hline
$a^{(1)}_n$ & $c^{V,(1)}_{(1)\,n}$ & $c^{V,(1)}_{(2)\,n}$ & 
$c^{V,(1)}_{(3)\,n}$ & $c^{V,(1)}_{(6)\,n}$ \\
\hline
$1$ & $2$ & $8/3$ & $4/3$ & $0$ \\
$\pi^2 \alpha$ & $- 8/27$ & $- 8/27$ & $- 8/27$ & $0$ \\
$\alpha$ & $- 1$ & $- 4/3$ & $- 2/3$ & $0$ \\
$\pi^2$ & $4/27$ & $8/27$ & $0$ & $8/27$ \\
$\psi^\prime(1/3)$ & $- 2/9$ & $- 4/9$ & $0$ & $- 4/9$ \\
$\psi^\prime(1/3) \alpha$ & $4/9$ & $4/9$ & $4/9$ & $0$ \\
\hline
\end{tabular}
\end{center}
\begin{center}
{Table $9$. RI${}^\prime$/SMOM coefficients of $C_F$ for one loop $V$ 
amplitudes.}
\end{center}
\end{table}}

\vspace{4cm}
{\begin{table}[hb]
\begin{center}
\begin{tabular}{|c||r|r|r|r|}
\hline
$a^{(21)}_n$ & $c^{V,(21)}_{(1)\,n}$ & $c^{V,(21)}_{(2)\,n}$ & 
$c^{V,(21)}_{(3)\,n}$ & $c^{V,(21)}_{(6)\,n}$ \\
\hline
$1$ & $- 58/9$ & $- 232/27$ & $- 116/27$ & $0$ \\
$\pi^2 \alpha$ & $160/243$ & $160/243$ & $160/243$ & $0$ \\
$\alpha$ & $20/9$ & $80/27$ & $40/27$ & $0$ \\
$\pi^2$ & $- 152/243$ & $- 304/243$ & $0$ & $- 208/243$ \\
$\psi^\prime(1/3)$ & $76/81$ & $152/81$ & $0$ & $104/81$ \\
$\psi^\prime(1/3) \alpha$ & $- 80/81$ & $- 80/81$ & $- 80/81$ & $0$ \\
\hline
\end{tabular}
\end{center}
\begin{center}
{Table $10$. RI${}^\prime$/SMOM coefficients of $C_F T_F \Nf$ for two loop $V$ 
amplitudes.}
\end{center}
\end{table}}

\clearpage

{\begin{table}[ht]
\begin{center}
\begin{tabular}{|c||r|r|r|r|}
\hline
$a^{(22)}_n$ & $c^{V,(22)}_{(1)\,n}$ & $c^{V,(22)}_{(2)\,n}$ & 
$c^{V,(22)}_{(3)\,n}$ & $c^{V,(22)}_{(6)\,n}$ \\
\hline
$1$ & $707/36$ & $707/27$ & $707/54$ & $0$ \\
$\pi^2 \alpha$ & $- 284/243$ & $- 446/243$ & $- 122/243$ & $2/9$ \\
$\pi^4 \alpha$ & $4/81$ & $4/81$ & $4/81$ & $- 4/81$ \\
$\zeta(3) \alpha$ & $0$ & $0$ & $0$ & $- 2/3$ \\
$\Sigma \alpha$ & $2/3$ & $2/3$ & $2/3$ & $0$ \\
$\alpha$ & $- 223/36$ & $- 223/27$ & $- 223/54$ & $0$ \\
$\pi^2 \alpha^2$ & $- 10/27$ & $- 10/27$ & $- 10/27$ & $2/27$ \\
$\alpha^2$ & $- 5/4$ & $- 5/3$ & $- 5/6$ & $0$ \\
$\pi^2 \alpha^3$ & $- 2/27$ & $- 2/27$ & $- 2/27$ & $0$ \\
$\alpha^3$ & $- 1/4$ & $- 1/3$ & $- 1/6$ & $0$ \\
$\pi^2$ & $- 236/243$ & $266/243$ & $- 82/27$ & $- 1672/243$ \\
$\pi^4$ & $- 10/81$ & $- 32/243$ & $- 28/243$ & $- 32/81$ \\
$\zeta(3)$ & $- 11/3$ & $- 2$ & $- 16/3$ & $- 28/3$ \\
$\Sigma$ & $- 2/3$ & $- 4/3$ & $0$ & $- 4/3$ \\
$s_2(\pi/6)$ & $10$ & $0$ & $20$ & $68$ \\
$s_2(\pi/6) \alpha$ & $- 6$ & $0$ & $- 12$ & $4$ \\
$s_2(\pi/2)$ & $20$ & $0$ & $- 40$ & $- 136$ \\
$s_2(\pi/2) \alpha$ & $12$ & $0$ & $24$ & $- 8$ \\
$s_3(\pi/6)$ & $- 50/3$ & $0$ & $- 100/3$ & $- 340/3$ \\
$s_3(\pi/6) \alpha$ & $10$ & $0$ & $20$ & $- 20/3$ \\
$s_3(\pi/2)$ & $40/3$ & $0$ & $80/3$ & $272/3$ \\
$s_3(\pi/2) \alpha$ & $- 8$ & $0$ & $- 16$ & $16/3$ \\
$\psi^\prime(1/3)$ & $118/81$ & $- 133/81$ & $41/9$ & $836/81$ \\
$\psi^\prime(1/3) \alpha$ & $142/81$ & $233/81$ & $61/81$ & $- 1/3$ \\
$\psi^\prime(1/3) \alpha^2$ & $5/9$ & $5/9$ & $5/9$ & $- 1/9$ \\
$\psi^\prime(1/3) \alpha^3$ & $1/9$ & $1/9$ & $1/9$ & $0$ \\
$\psi^\prime(1/3) \pi^2 $ & $8/27$ & $32/81$ & $16/81$ & $16/27$ \\
$(\psi^\prime(1/3))^2$ & $- 2/9$ & $- 8/27$ & $- 4/27$ & $- 4/9$ \\
$\psi^{\prime\prime\prime}(1/3)$ & $1/108$ & $0$ & $1/54$ & $2/27$ \\
$\psi^{\prime\prime\prime}(1/3) \alpha$ & $- 1/54$ & $- 1/54$ & $- 1/54$ & $1/54$ \\
$\pi^3 \alpha/\sqrt{3}$ & $29/648$ & $0$ & $29/324$ & $- 29/972$ \\
$\pi^3/\sqrt{3}$ & $- 145/1944$ & $0$ & $- 145/972$ & $- 493/972$ \\
$\pi \ln(3) \alpha/\sqrt{3}$ & $1/2$ & $0$ & $1$ & $- 1/3$ \\
$\pi \ln(3)/\sqrt{3}$ & $- 5/6$ & $0$ & $- 5/3$ & $- 17/3$ \\
$\pi (\ln(3))^2 \alpha/\sqrt{3}$ & $- 1/24$ & $0$ & $- 1/12$ & $1/36$ \\
$\pi (\ln(3))^2/\sqrt{3}$ & $5/72$ & $0$ & $5/36$ & $17/36$ \\
\hline
\end{tabular}
\end{center}
\begin{center}
{Table $11$. RI${}^\prime$/SMOM coefficients of $C_F C_A$ for two loop $V$ 
amplitudes.}
\end{center}
\end{table}}

\clearpage

{\begin{table}[ht]
\begin{center}
\begin{tabular}{|c||r|r|r|r|}
\hline
$a^{(23)}_n$ & $c^{V,(23)}_{(1)\,n}$ & $c^{V,(23)}_{(2)\,n}$ & 
$c^{V,(23)}_{(3)\,n}$ & $c^{V,(23)}_{(6)\,n}$ \\
\hline
$1$ & $- 7/2$ & $- 14/3$ & $- 7/3$ & $0$ \\
$\pi^2 \alpha$ & $- 68/27$ & $- 28/27$ & $- 4$ & $- 8/27$ \\
$\pi^4 \alpha$ & $16/81$ & $16/81$ & $16/81$ & $0$ \\
$\zeta(3) \alpha$ & $4/3$ & $4/3$ & $8/3$ & $0$ \\
$\Sigma \alpha$ & $4/3$ & $4/3$ & $4/3$ & $0$ \\
$\alpha$ & $2$ & $8/3$ & $4/3$ & $0$ \\
$\pi^2 \alpha^2$ & $0$ & $0$ & $0$ & $- 4/27$ \\
$\pi^4 \alpha^2$ & $0$ & $0$ & $0$ & $0$ \\
$\alpha^2$ & $0$ & $0$ & $0$ & $0$ \\
$\pi^2$ & $122/27$ & $88/27$ & $52/9$ & $179/9$ \\
$\pi^4$ & $- 20/81$ & $- 128/243$ & $8/243$ & $8/27$ \\
$\zeta(3)$ & $- 28/3$ & $- 16$ & $- 8/3$ & $8$ \\
$\Sigma$ & $- 2/3$ & $- 4/3$ & $0$ & $- 4/3$ \\
$s_2(\pi/6)$ & $8$ & $48$ & $- 32$ & $- 144$ \\
$s_2(\pi/6) \alpha$ & $16$ & $0$ & $32$ & $0$ \\
$s_2(\pi/2)$ & $- 16$ & $- 96$ & $64$ & $288$ \\
$s_2(\pi/2) \alpha$ & $- 32$ & $0$ & $- 64$ & $0$ \\
$s_3(\pi/6)$ & $- 40/3$ & $- 80$ & $160/3$ & $240$ \\
$s_3(\pi/6) \alpha$ & $- 80/3$ & $0$ & $- 160/3$ & $0$ \\
$s_3(\pi/2)$ & $32/3$ & $64$ & $- 128/3$ & $- 192$ \\
$s_3(\pi/2) \alpha$ & $64/3$ & $0$ & $128/3$ & $0$ \\
$\psi^\prime(1/3)$ & $- 61/9$ & $- 44/9$ & $- 26/3$ & $- 86/3$ \\
$\psi^\prime(1/3) \pi^2 \alpha$ & $0$ & $0$ & $0$ & $0$ \\
$\psi^\prime(1/3) \alpha$ & $34/9$ & $14/9$ & $6$ & $4/9$ \\
$\psi^\prime(1/3) \pi^2 \alpha^2 $ & $0$ & $0$ & $0$ & $0$ \\
$\psi^\prime(1/3) \alpha^2 $ & $0$ & $0$ & $0$ & $2/9$ \\
$\psi^\prime(1/3) \pi^2 $ & $- 16/27$ & $- 64/81$ & $- 32/81$ & $- 32/27$ \\
$(\psi^\prime(1/3))^2$ & $4/9$ & $16/27$ & $8/27$ & $8/9$ \\
$(\psi^\prime(1/3))^2 \alpha$ & $0$ & $0$ & $0$ & $0$ \\
$(\psi^\prime(1/3))^2 \alpha^2$ & $0$ & $0$ & $0$ & $0$ \\
$\psi^{\prime\prime\prime}(1/3)$ & $1/6$ & $8/27$ & $1/27$ & $1/27$ \\
$\psi^{\prime\prime\prime}(1/3) \alpha$ & $- 2/27$ & $- 2/27$ & $- 2/27$ & $0$ \\
$\pi^3 \alpha/\sqrt{3}$ & $- 29/243$ & $0$ & $- 58/243$ & $0$ \\
$\pi^3/\sqrt{3}$ & $- 29/486$ & $- 29/81$ & $58/243$ & $29/27$ \\
$\pi \ln(3) \alpha/\sqrt{3}$ & $- 4/3$ & $0$ & $- 8/3$ & $0$ \\
$\pi \ln(3)/\sqrt{3}$ & $- 2/3$ & $- 4$ & $8/3$ & $12$ \\
$\pi (\ln(3))^2 \alpha/\sqrt{3}$ & $1/9$ & $0$ & $2/9$ & $0$ \\
$\pi (\ln(3))^2/\sqrt{3}$ & $1/18$ & $1/3$ & $- 2/9$ & $- 1$ \\
\hline
\end{tabular}
\end{center}
\begin{center}
{Table $12$. RI${}^\prime$/SMOM coefficients of $C_F^2$ for two loop $V$ 
amplitudes.}
\end{center}
\end{table}}

\clearpage 

{\begin{table}[ht]
\begin{center}
\begin{tabular}{|c||r|r|r|r|r|r|}
\hline
$a^{(1)}_n$ & $\left. c^{T,(1)}_{(1)\,n} \right|_{\MSbars}$ & 
$\left. c^{T,(1)}_{(2)\,n} \right|_{\MSbars}$ & 
$\left. c^{T,(1)}_{(3)\,n} \right|_{\MSbars}$ & 
$\left. c^{T,(1)}_{(4)\,n} \right|_{\MSbars}$ & 
$\left. c^{T,(1)}_{(7)\,n} \right|_{\MSbars}$ & 
$\left. c^{T,(1)}_{(8)\,n} \right|_{\MSbars}$ \\
\hline
$1$ & $2$ & $0$ & $4/3$ & $2/3$ & $8/3$ & $0$ \\
$\pi^2 \alpha$ & $- 10/27$ & $- 4/27$ & $- 8/27$ & $- 4/27$ & $- 16/27$ & $4/27$ \\
$\alpha$ & $- 2$ & $0$ & $- 4/3$ & $-2/3$ & $- 8/3$ & $0$ \\
$\pi^2$ & $10/27$ & $- 4/9$ & $8/27$ & $4/27$ & $16/27$ & $4/9$ \\
$\psi^\prime(1/3)$ & $- 5/9$ & $2/3$ & $- 4/9$ & $- 2/9$ & $- 8/9$ & $- 2/3$ \\
$\psi^\prime(1/3) \alpha$ & $5/9$ & $2/9$ & $4/9$ & $2/9$ & $8/9$ & $- 2/9$ \\
\hline
\end{tabular}
\end{center}
\begin{center}
{Table $13$. $\MSbar$ coefficients of $C_F$ for one loop $T$ amplitudes.}
\end{center}
\end{table}}

\vspace{4cm}
{\begin{table}[hb]
\begin{center}
\begin{tabular}{|c||r|r|r|r|r|r|}
\hline
$a^{(21)}_n$ & $\left. c^{T,(21)}_{(1)\,n} \right|_{\MSbars}$ & 
$\left. c^{T,(21)}_{(2)\,n} \right|_{\MSbars}$ & 
$\left. c^{T,(21)}_{(3)\,n} \right|_{\MSbars}$ & 
$\left. c^{T,(21)}_{(4)\,n} \right|_{\MSbars}$ & 
$\left. c^{T,(21)}_{(7)\,n} \right|_{\MSbars}$ & 
$\left. c^{T,(21)}_{(8)\,n} \right|_{\MSbars}$ \\ 
\hline
$1$ & $- 182/27$ & $0$ & $- 80/27$ & $- 40/27$ & $- 160/27$ & $0$ \\
$\pi^2 \alpha$ & $0$ & $0$ & $0$ & $0$ & $0$ & $0$ \\
$\alpha$ & $0$ & $0$ & $0$ & $0$ & $0$ & $0$ \\
$\pi^2$ & $- 200/243$ & $128/81$ & $- 160/243$ & $- 80/243$ & $- 320/243$ & $- 128/81$ \\
$\psi^\prime(1/3)$ & $100/81$ & $- 64/27$ & $80/81$ & $40/81$ & $160/81$ & $64/27$ \\
$\psi^\prime(1/3) \alpha$ & $0$ & $0$ & $0$ & $0$ & $0$ & $0$ \\
\hline
\end{tabular}
\end{center}
\begin{center}
{Table $14$. $\MSbar$ coefficients of $C_F T_F \Nf$ for two loop $T$ 
amplitudes.}
\end{center}
\end{table}}

\clearpage

{\begin{table}[ht]
\begin{center}
\begin{tabular}{|c||r|r|r|r|r|r|}
\hline
$a^{(22)}_n$ & $\left. c^{T,(22)}_{(1)\,n} \right|_{\MSbars}$ & 
$\left. c^{T,(22)}_{(2)\,n} \right|_{\MSbars}$ & 
$\left. c^{T,(22)}_{(3)\,n} \right|_{\MSbars}$ & 
$\left. c^{T,(22)}_{(4)\,n} \right|_{\MSbars}$ & 
$\left. c^{T,(22)}_{(7)\,n} \right|_{\MSbars}$ & 
$\left. c^{T,(22)}_{(8)\,n} \right|_{\MSbars}$ \\ 
\hline 
$1$ & $7097/216$ & $0$ & $205/27$ & $205/54$ & $410/27$ & $0$ \\ 
$\pi^2 \alpha$ & $1/27$ & $16/27$ & $- 28/27$ & $- 14/27$ & $- 56/27$ & $- 16/27$ \\ 
$\pi^4 \alpha$ & $5/81$ & $2/27$ & $4/81$ & $2/81$ & $8/81$ & $- 2/27$ \\ 
$\zeta(3) \alpha$ & $3$ & $2/3$ & $0$ & $0$ & $0$ & $- 2/3$ \\
$\Sigma \alpha$ & $5/6$ & $1/3$ & $2/3$ & $1/3$ & $4/3$ & $- 1/3$ \\
$\alpha$ & $- 10$ & $0$ & $- 14/3$ & $- 7/3$ & $- 28/3$ & $0$ \\
$\pi^2 \alpha^2$ & $- 5/18$ & $- 5/27$ & $- 2/9$ & $- 1/9$ & $- 4/9$ & $5/27$ \\
$\alpha^2$ & $- 15/8$ & $0$ & $- 1$ & $- 1/2$ & $- 2$ & $0$ \\
$\pi^2 \alpha^3$ & $0$ & $0$ & $0$ & $0$ & $0$ & $0$ \\
$\alpha^3$ & $0$ & $0$ & $0$ & $0$ & $0$ & $0$ \\
$\pi^2$ & $- 919/486$ & $503/81$ & $1346/243$ & $673/243$ & $2692/243$ & $- 503/81$ \\
$\pi^4$ & $- 1/27$ & $2/3$ & $52/243$ & $26/243$ & $104/243$ & $- 2/3$ \\
$\zeta(3)$ & $- 9$ & $26/3$ & $28/3$ & $14/3$ & $56/3$ & $- 26/3$ \\
$\Sigma$ & $- 7/6$ & $5/3$ & $- 2/3$ & $- 1/3$ & $- 4/3$ & $- 5/3$ \\
$s_2(\pi/6)$ & $24$ & $- 112$ & $- 80$ & $- 40$ & $- 160$ & $112$ \\
$s_2(\pi/6) \alpha$ & $- 12$ & $- 16$ & $0$ & $0$ & $0$ & $16$ \\
$s_2(\pi/2)$ & $- 48$ & $224$ & $160$ & $80$ & $320$ & $- 224$ \\
$s_2(\pi/2) \alpha$ & $24$ & $32$ & $0$ & $0$ & $0$ & $- 32$ \\
$s_3(\pi/6)$ & $- 40$ & $560/3$ & $400/3$ & $200/3$ & $800/3$ & $- 560/3$ \\
$s_3(\pi/6) \alpha$ & $20$ & $80/3$ & $0$ & $0$ & $0$ & $- 80/3$ \\
$s_3(\pi/2)$ & $32$ & $- 448/3$ & $- 320/3$ & $- 160/3$ & $- 640/3$ & $448/3$ \\
$s_3(\pi/2) \alpha$ & $- 16$ & $- 64/3$ & $0$ & $0$ & $0$ & $64/3$ \\
$\psi^\prime(1/3)$ & $919/324$ & $- 503/54$ & $- 673/81$ & $- 673/162$ & $- 1346/81$ & $503/54$ \\
$\psi^\prime(1/3) \alpha$ & $- 1/18$ & $- 8/9$ & $14/9$ & $7/9$ & $28/9$ & $8/9$ \\
$\psi^\prime(1/3) \alpha^2$ & $5/12$ & $5/18$ & $1/3$ & $1/6$ & $2/3$ & $- 5/18$ \\
$\psi^\prime(1/3) \alpha^3$ & $0$ & $0$ & $0$ & $0$ & $0$ & $0$ \\
$\psi^\prime(1/3) \pi^2 $ & $8/27$ & $- 16/27$ & $32/81$ & $16/81$ & $64/81$ & $16/27$ \\
$(\psi^\prime(1/3))^2$ & $- 2/9$ & $4/9$ & $- 8/27$ & $- 4/27$ & $- 16/27$ & $- 4/9$ \\
$\psi^{\prime\prime\prime}(1/3)$ & $- 5/216$ & $- 19/108$ & $- 7/54$ & $- 7/108$ & $- 7/27$ & $19/108$ \\
$\psi^{\prime\prime\prime}(1/3) \alpha$ & $- 5/216$ & $- 1/36$ & $- 1/54$ & $- 1/108$ & $- 1/27$ & $1/36$ \\
$\pi^3 \alpha/\sqrt{3}$ & $29/324$ & $29/243$ & $0$ & $0$ & $0$ & $- 29/243$ \\
$\pi^3/\sqrt{3}$ & $- 29/162$ & $203/243$ & $145/243$ & $145/486$ & $290/243$ & $- 203/243$ \\
$\pi \ln(3) \alpha/\sqrt{3}$ & $1$ & $4/3$ & $0$ & $0$ & $0$ & $- 4/3$ \\
$\pi \ln(3)/\sqrt{3}$ & $- 2$ & $28/3$ & $20/3$ & $10/3$ & $40/3$ & $- 28/3$ \\
$\pi (\ln(3))^2 \alpha/\sqrt{3}$ & $- 1/12$ & $- 1/9$ & $0$ & $0$ & $0$ & $1/9$ \\
$\pi (\ln(3))^2/\sqrt{3}$ & $1/6$ & $- 7/9$ & $- 5/9$ & $- 5/18$ & $- 10/9$ & $7/9$ \\
\hline
\end{tabular}
\end{center}
\begin{center}
{Table $15$. $\MSbar$ coefficients of $C_F C_A$ for two loop $T$ amplitudes.}
\end{center}
\end{table}}

\clearpage

{\begin{table}[ht]
\begin{center}
\begin{tabular}{|c||r|r|r|r|r|r|}
\hline
$a^{(23)}_n$ & $\left. c^{T,(23)}_{(1)\,n} \right|_{\MSbars}$ & 
$\left. c^{T,(23)}_{(2)\,n} \right|_{\MSbars}$ & 
$\left. c^{T,(23)}_{(3)\,n} \right|_{\MSbars}$ & 
$\left. c^{T,(23)}_{(4)\,n} \right|_{\MSbars}$ &
$\left. c^{T,(23)}_{(7)\,n} \right|_{\MSbars}$ & 
$\left. c^{T,(23)}_{(8)\,n} \right|_{\MSbars}$ \\
\hline
$1$ & $- 79/3$ & $0$ & $- 4$ & $- 2$ & $- 8$ & $0$ \\
$\pi^2 \alpha$ & $- 110/27$ & $- 100/27$ & $- 16/27$ & $- 8/27$ & $- 32/27$ & $100/27$ \\
$\pi^4 \alpha$ & $20/81$ & $8/81$ & $16/81$ & $8/81$ & $32/81$ & $- 8/81$ \\
$\zeta(3) \alpha$ & $8/3$ & $8/3$ & $0$ & $0$ & $0$ & $- 8/3$ \\
$\Sigma \alpha$ & $5/3$ & $2/3$ & $4/3$ & $2/3$ & $8/3$ & $- 2/3$ \\
$\alpha$ & $4$ & $0$ & $16/3$ & $8/3$ & $32/3$ & $0$ \\
$\pi^2 \alpha^2$ & $- 10/27$ & $0$ & $- 8/27$ & $- 4/27$ & $- 16/27$ & $0$ \\
$\pi^4 \alpha^2$ & $0$ & $0$ & $0$ & $0$ & $0$ & $0$ \\
$\alpha^2$ & $- 1$ & $0$ & $- 4/3$ & $- 2/3$ & $- 8/3$ & $0$ \\
$\pi^2$ & $128/9$ & $- 148/9$ & $56/27$ & $28/27$ & $112/27$ & $148/9$ \\
$\pi^4$ & $- 32/81$ & $- 16/81$ & $- 128/243$ & $- 64/243$ & $- 256/243$ & $16/81$ \\
$\zeta(3)$ & $20/3$ & $8/3$ & $- 32/3$ & $- 16/3$ & $- 64/3$ & $- 8/3$ \\
$\Sigma$ & $- 5/3$ & $2$ & $- 4/3$ & $- 2/3$ & $- 8/3$ & $- 2$ \\
$s_2(\pi/6)$ & $- 88$ & $128$ & $64$ & $32$ & $128$ & $- 128$ \\
$s_2(\pi/6) \alpha$ & $32$ & $32$ & $0$ & $0$ & $0$ & $- 32$ \\
$s_2(\pi/2)$ & $176$ & $- 256$ & $- 128$ & $- 64$ & $- 256$ & $256$ \\
$s_2(\pi/2) \alpha$ & $- 64$ & $- 64$ & $0$ & $0$ & $0$ & $64$ \\
$s_3(\pi/6)$ & $440/3$ & $- 640/3$ & $- 320/3$ & $- 160/3$ & $- 640/3$ & $640/3$ \\
$s_3(\pi/6) \alpha$ & $- 160/3$ & $- 160/3$ & $0$ & $0$ & $0$ & $160/3$ \\
$s_3(\pi/2)$ & $- 352/3$ & $512/3$ & $256/3$ & $128/3$ & $512/3$ & $- 512/3$ \\
$s_3(\pi/2) \alpha$ & $128/3$ & $128/3$ & $0$ & $0$ & $0$ & $- 128/3$ \\
$\psi^\prime(1/3)$ & $- 64/3$ & $74/3$ & $- 28/9$ & $- 14/9$ & $- 56/9$ & $- 74/3$ \\
$\psi^\prime(1/3) \pi^2 \alpha$ & $0$ & $0$ & $0$ & $0$ & $0$ & $0$ \\
$\psi^\prime(1/3) \alpha$ & $55/9$ & $50/9$ & $8/9$ & $4/9$ & $16/9$ & $- 50/9$ \\
$\psi^\prime(1/3) \pi^2 \alpha^2 $ & $0$ & $0$ & $0$ & $0$ & $0$ & $0$ \\
$\psi^\prime(1/3) \alpha^2 $ & $5/9$ & $0$ & $4/9$ & $2/9$ & $8/9$ & $0$ \\
$\psi^\prime(1/3) \pi^2 $ & $- 16/27$ & $32/27$ & $- 64/81$ & $- 32/81$ & $- 128/81$ & $- 32/27$ \\
$(\psi^\prime(1/3))^2$ & $4/9$ & $- 8/9$ & $16/27$ & $8/27$ & $32/27$ & $8/9$ \\
$(\psi^\prime(1/3))^2 \alpha$ & $0$ & $0$ & $0$ & $0$ & $0$ & $0$ \\
$(\psi^\prime(1/3))^2 \alpha^2$ & $0$ & $0$ & $0$ & $0$ & $0$ & $0$ \\
$\psi^{\prime\prime\prime}(1/3)$ & $2/9$ & $- 2/27$ & $8/27$ & $4/27$ & $16/27$ & $2/27$ \\
$\psi^{\prime\prime\prime}(1/3) \alpha$ & $- 5/54$ & $- 1/27$ & $- 2/27$ & $- 1/27$ & $- 4/27$ & $1/27$ \\
$\pi^3 \alpha/\sqrt{3}$ & $- 58/243$ & $- 58/243$ & $0$ & $0$ & $0$ & $58/243$ \\
$\pi^3/\sqrt{3}$ & $319/486$ & $- 232/243$ & $- 116/243$ & $- 58/243$ & $- 232/243$ & $232/243$ \\
$\pi \ln(3) \alpha/\sqrt{3}$ & $- 8/3$ & $- 8/3$ & $0$ & $0$ & $0$ & $8/3$ \\
$\pi \ln(3)/\sqrt{3}$ & $22/3$ & $- 32/3$ & $- 16/3$ & $- 8/3$ & $- 32/3$ & $32/3$ \\
$\pi (\ln(3))^2 \alpha/\sqrt{3}$ & $2/9$ & $2/9$ & $0$ & $0$ & $0$ & $- 2/9$ \\
$\pi (\ln(3))^2/\sqrt{3}$ & $- 11/18$ & $8/9$ & $4/9$ & $2/9$ & $8/9$ & $- 8/9$ \\
\hline
\end{tabular}
\end{center}
\begin{center}
{Table $16$. $\MSbar$ coefficients of $C_F^2$ for two loop $T$ amplitudes.}
\end{center}
\end{table}}

\clearpage 

{\begin{table}[ht]
\begin{center}
\begin{tabular}{|c||r|r|r|r|r|r|}
\hline
$a^{(1)}_n$ & $c^{T,(1)}_{(1)\,n}$ & $c^{T,(1)}_{(2)\,n}$ & 
$c^{T,(1)}_{(3)\,n}$ & $c^{T,(1)}_{(4)\,n}$ & $c^{T,(1)}_{(7)\,n}$ & 
$c^{T,(1)}_{(8)\,n}$ \\
\hline
$1$ & $2/3$ & $0$ & $4/3$ & $2/3$ & $8/3$ & $0$ \\
$\pi^2 \alpha$ & $- 4/27$ & $- 4/27$ & $- 8/27$ & $- 4/27$ & $- 16/27$ & $4/27$ \\
$\alpha$ & $- 2/3$ & $0$ & $- 4/3$ & $-2/3$ & $- 8/3$ & $0$ \\
$\pi^2$ & $4/27$ & $- 4/9$ & $8/27$ & $4/27$ & $16/27$ & $4/9$ \\
$\psi^\prime(1/3)$ & $- 2/9$ & $2/3$ & $- 4/9$ & $- 2/9$ & $- 8/9$ & $- 2/3$ \\
$\psi^\prime(1/3) \alpha$ & $2/9$ & $2/9$ & $4/9$ & $2/9$ & $8/9$ & $- 2/9$ \\
\hline
\end{tabular}
\end{center}
\begin{center}
{Table $17$. RI${}^\prime$/SMOM coefficients of $C_F$ for one loop $T$ 
amplitudes.}
\end{center}
\end{table}}

\vspace{4cm}
{\begin{table}[hb]
\begin{center}
\begin{tabular}{|c||r|r|r|r|r|r|}
\hline
$a^{(21)}_n$ & $c^{T,(21)}_{(1)\,n}$ & $c^{T,(21)}_{(2)\,n}$ & 
$c^{T,(21)}_{(3)\,n}$ & $c^{T,(21)}_{(4)\,n}$ & $c^{T,(21)}_{(7)\,n}$ & 
$c^{T,(21)}_{(8)\,n}$ \\ 
\hline
$1$ & $- 40/27$ & $0$ & $- 80/27$ & $- 40/27$ & $- 160/27$ & $0$ \\
$\pi^2 \alpha$ & $80/243$ & $80/243$ & $160/243$ & $80/243$ & $320/243$ & $- 80/243$ \\
$\alpha$ & $40/27$ & $0$ & $80/27$ & $40/27$ & $160/27$ & $0$ \\
$\pi^2$ & $- 80/243$ & $128/81$ & $- 160/243$ & $- 80/243$ & $- 320/243$ & $- 128/81$ \\
$\psi^\prime(1/3)$ & $40/81$ & $- 64/27$ & $80/81$ & $40/81$ & $160/81$ & $64/27$ \\
$\psi^\prime(1/3) \alpha$ & $- 40/81$ & $- 40/81$ & $- 80/81$ & $- 40/81$ & $- 160/81$ & $40/81$ \\
\hline
\end{tabular}
\end{center}
\begin{center}
{Table $18$. RI${}^\prime$/SMOM coefficients of $C_F T_F \Nf$ for two loop $T$ 
amplitudes.}
\end{center}
\end{table}}

\clearpage

{\begin{table}[ht]
\begin{center}
\begin{tabular}{|c||r|r|r|r|r|r|}
\hline
$a^{(22)}_n$ & $c^{T,(22)}_{(1)\,n}$ & $c^{T,(22)}_{(2)\,n}$ & 
$c^{T,(22)}_{(3)\,n}$ & $c^{T,(22)}_{(4)\,n}$ & $c^{T,(22)}_{(7)\,n}$ & 
$c^{T,(22)}_{(8)\,n}$ \\ 
\hline 
$1$ & $205/54$ & $0$ & $205/27$ & $205/54$ & $410/27$ & $0$ \\ 
$\pi^2 \alpha$ & $- 223/243$ & $47/243$ & $- 446/243$ & $- 223/243$ & $- 892/243$ & $- 47/243$ \\ 
$\pi^4 \alpha$ & $2/81$ & $2/27$ & $4/81$ & $2/81$ & $8/81$ & $- 2/27$ \\ 
$\zeta(3) \alpha$ & $0$ & $2/3$ & $0$ & $0$ & $0$ & $- 2/3$ \\
$\Sigma \alpha$ & $1/3$ & $1/3$ & $2/3$ & $1/3$ & $4/3$ & $- 1/3$ \\
$\alpha$ & $- 223/54$ & $0$ & $- 223/27$ & $- 223/54$ & $- 446/27$ & $0$ \\
$\pi^2 \alpha^2$ & $- 5/27$ & $- 7/27$ & $- 10/27$ & $- 5/27$ & $- 20/27$ & $7/27$ \\
$\alpha^2$ & $- 5/6$ & $0$ & $- 5/3$ & $- 5/6$ & $- 10/3$ & $0$ \\
$\pi^2 \alpha^3$ & $- 1/27$ & $- 1/27$ & $- 2/27$ & $- 1/27$ & $- 4/27$ & $1/27$ \\
$\alpha^3$ & $- 1/6$ & $0$ & $- 1/3$ & $- 1/6$ & $- 2/3$ & $0$ \\
$\pi^2$ & $673/243$ & $503/81$ & $1346/243$ & $673/243$ & $2692/243$ & $- 503/81$ \\
$\pi^4$ & $26/243$ & $2/3$ & $52/243$ & $26/243$ & $104/243$ & $- 2/3$ \\
$\zeta(3)$ & $14/3$ & $26/3$ & $28/3$ & $14/3$ & $56/3$ & $- 26/3$ \\
$\Sigma$ & $- 1/3$ & $5/3$ & $- 2/3$ & $- 1/3$ & $- 4/3$ & $- 5/3$ \\
$s_2(\pi/6)$ & $- 40$ & $- 112$ & $- 80$ & $- 40$ & $- 160$ & $112$ \\
$s_2(\pi/6) \alpha$ & $0$ & $- 16$ & $0$ & $0$ & $0$ & $16$ \\
$s_2(\pi/2)$ & $80$ & $224$ & $160$ & $80$ & $320$ & $- 224$ \\
$s_2(\pi/2) \alpha$ & $0$ & $32$ & $0$ & $0$ & $0$ & $- 32$ \\
$s_3(\pi/6)$ & $200/3$ & $560/3$ & $400/3$ & $200/3$ & $800/3$ & $- 560/3$ \\
$s_3(\pi/6) \alpha$ & $0$ & $80/3$ & $0$ & $0$ & $0$ & $- 80/3$ \\
$s_3(\pi/2)$ & $- 160/3$ & $- 448/3$ & $- 320/3$ & $- 160/3$ & $- 640/3$ & $448/3$ \\
$s_3(\pi/2) \alpha$ & $0$ & $- 64/3$ & $0$ & $0$ & $0$ & $64/3$ \\
$\psi^\prime(1/3)$ & $- 673/162$ & $- 503/54$ & $- 673/81$ & $- 673/162$ & $- 1346/81$ & $503/54$ \\
$\psi^\prime(1/3) \alpha$ & $223/162$ & $- 47/162$ & $223/81$ & $223/162$ & $446/819$ & $47/162$ \\
$\psi^\prime(1/3) \alpha^2$ & $5/18$ & $7/18$ & $5/9$ & $5/18$ & $10/9$ & $- 7/18$ \\
$\psi^\prime(1/3) \alpha^3$ & $1/18$ & $1/18$ & $1/9$ & $1/18$ & $2/9$ & $- 1/18$ \\
$\psi^\prime(1/3) \pi^2 $ & $16/81$ & $- 16/27$ & $32/81$ & $16/81$ & $64/81$ & $16/27$ \\
$(\psi^\prime(1/3))^2$ & $- 4/27$ & $4/9$ & $- 8/27$ & $- 4/27$ & $- 16/27$ & $- 4/9$ \\
$\psi^{\prime\prime\prime}(1/3)$ & $- 7/108$ & $- 19/108$ & $- 7/54$ & $- 7/108$ & $- 7/27$ & $19/108$ \\
$\psi^{\prime\prime\prime}(1/3) \alpha$ & $- 1/108$ & $- 1/36$ & $- 1/54$ & $- 1/108$ & $- 1/27$ & $1/36$ \\
$\pi^3 \alpha/\sqrt{3}$ & $0$ & $29/243$ & $0$ & $0$ & $0$ & $- 29/243$ \\
$\pi^3/\sqrt{3}$ & $145/486$ & $203/243$ & $145/243$ & $145/486$ & $290/243$ & $- 203/243$ \\
$\pi \ln(3) \alpha/\sqrt{3}$ & $0$ & $4/3$ & $0$ & $0$ & $0$ & $- 4/3$ \\
$\pi \ln(3)/\sqrt{3}$ & $10/3$ & $28/3$ & $20/3$ & $10/3$ & $40/3$ & $- 28/3$ \\
$\pi (\ln(3))^2 \alpha/\sqrt{3}$ & $0$ & $- 1/9$ & $0$ & $0$ & $0$ & $1/9$ \\
$\pi (\ln(3))^2/\sqrt{3}$ & $- 5/18$ & $- 7/9$ & $- 5/9$ & $- 5/18$ & $- 10/9$ & $7/9$ \\
\hline
\end{tabular}
\end{center}
\begin{center}
{Table $19$. RI${}^\prime$/SMOM coefficients of $C_F C_A$ for two loop $T$ 
amplitudes.}
\end{center}
\end{table}}

\clearpage

{\begin{table}[ht]
\begin{center}
\begin{tabular}{|c||r|r|r|r|r|r|}
\hline
$a^{(23)}_n$ & $c^{T,(23)}_{(1)\,n}$ & $c^{T,(23)}_{(2)\,n}$ & 
$c^{T,(23)}_{(3)\,n}$ & $c^{T,(23)}_{(4)\,n}$ & $c^{T,(23)}_{(7)\,n}$ & 
$c^{T,(23)}_{(8)\,n}$ \\
\hline
$1$ & $- 10/9$ & $0$ & $- 20/9$ & $- 10/9$ & $- 40/9$ & $0$ \\
$\pi^2 \alpha$ & $- 80/81$ & $- 268/81$ & $- 160/81$ & $- 80/81$ & $- 320/81$ & $268/81$ \\
$\pi^4 \alpha$ & $8/243$ & $40/2431$ & $16/243$ & $8/243$ & $32/243$ & $- 40/243$ \\
$\zeta(3) \alpha$ & $0$ & $8/3$ & $0$ & $0$ & $0$ & $- 8/3$ \\
$\Sigma \alpha$ & $2/3$ & $2/3$ & $4/3$ & $2/3$ & $8/3$ & $- 2/3$ \\
$\alpha$ & $8/9$ & $0$ & $16/81$ & $8/9$ & $32/9$ & $0$ \\
$\pi^2 \alpha^2$ & $16/81$ & $16/81$ & $32/81$ & $16/81$ & $64/81$ & $- 16/81$ \\
$\pi^4 \alpha^2$ & $8/243$ & $8/243$ & $16/243$ & $8/243$ & $32/243$ & $- 8/243$ \\
$\alpha^2$ & $2/9$ & $0$ & $4/9$ & $2/9$ & $8/9$ & $0$ \\
$\pi^2$ & $112/81$ & $- 460/27$ & $224/81$ & $26/27$ & $448/81$ & $460/27$ \\
$\pi^4$ & $- 56/243$ & $- 8/27$ & $- 112/243$ & $- 56/243$ & $- 224/243$ & $8/27$ \\
$\zeta(3)$ & $- 16/3$ & $8/3$ & $- 32/3$ & $- 16/3$ & $- 64/3$ & $- 8/3$ \\
$\Sigma$ & $- 2/3$ & $2$ & $- 4/3$ & $- 2/3$ & $- 8/3$ & $- 2$ \\
$s_2(\pi/6)$ & $32$ & $128$ & $64$ & $32$ & $128$ & $- 128$ \\
$s_2(\pi/6) \alpha$ & $0$ & $32$ & $0$ & $0$ & $0$ & $- 32$ \\
$s_2(\pi/2)$ & $- 64$ & $- 256$ & $- 128$ & $- 64$ & $- 256$ & $256$ \\
$s_2(\pi/2) \alpha$ & $0$ & $- 64$ & $0$ & $0$ & $0$ & $64$ \\
$s_3(\pi/6)$ & $- 160/3$ & $- 640/3$ & $- 320/3$ & $- 160/3$ & $- 640/3$ & $640/3$ \\
$s_3(\pi/6) \alpha$ & $0$ & $- 160/3$ & $0$ & $0$ & $0$ & $160/3$ \\
$s_3(\pi/2)$ & $128/3$ & $512/3$ & $256/3$ & $128/3$ & $512/3$ & $- 512/3$ \\
$s_3(\pi/2) \alpha$ & $0$ & $128/3$ & $0$ & $0$ & $0$ & $- 128/3$ \\
$\psi^\prime(1/3)$ & $- 56/27$ & $230/9$ & $- 112/27$ & $- 56/27$ & $- 224/27$ & $- 230/9$ \\
$\psi^\prime(1/3) \pi^2 \alpha$ & $16/81$ & $- 16/81$ & $32/81$ & $16/81$ & $64/81$ & $16/81$ \\
$\psi^\prime(1/3) \alpha$ & $40/27$ & $134/27$ & $80/27$ & $40/27$ & $160/27$ & $- 134/27$ \\
$\psi^\prime(1/3) \pi^2 \alpha^2 $ & $- 8/81$ & $- 8/81$ & $- 16/81$ & $- 8/81$ & $- 32/81$ & $8/81$ \\
$\psi^\prime(1/3) \alpha^2 $ & $- 8/27$ & $- 8/27$ & $- 16/27$ & $- 8/27$ & $- 32/27$ & $8/27$ \\
$\psi^\prime(1/3) \pi^2 $ & $- 40/81$ & $40/27$ & $- 80/81$ & $- 40/81$ & $- 160/81$ & $- 40/27$ \\
$(\psi^\prime(1/3))^2$ & $10/27$ & $- 10/9$ & $20/27$ & $10/27$ & $40/27$ & $10/9$ \\
$(\psi^\prime(1/3))^2 \alpha$ & $- 4/27$ & $4/27$ & $- 8/27$ & $- 4/27$ & $- 16/27$ & $- 4/27$ \\
$(\psi^\prime(1/3))^2 \alpha^2$ & $2/27$ & $2/27$ & $4/27$ & $2/27$ & $8/27$ & $- 2/27$ \\
$\psi^{\prime\prime\prime}(1/3)$ & $4/27$ & $- 2/27$ & $8/27$ & $4/27$ & $16/27$ & $2/27$ \\
$\psi^{\prime\prime\prime}(1/3) \alpha$ & $- 1/27$ & $- 1/27$ & $- 2/27$ & $- 1/27$ & $- 4/27$ & $1/27$ \\
$\pi^3 \alpha/\sqrt{3}$ & $0$ & $- 58/243$ & $0$ & $0$ & $0$ & $58/243$ \\
$\pi^3/\sqrt{3}$ & $- 58/243$ & $- 232/243$ & $- 116/243$ & $- 58/243$ & $- 232/243$ & $232/243$ \\
$\pi \ln(3) \alpha/\sqrt{3}$ & $0$ & $- 8/3$ & $0$ & $0$ & $0$ & $8/3$ \\
$\pi \ln(3)/\sqrt{3}$ & $- 8/3$ & $- 32/3$ & $- 16/3$ & $- 8/3$ & $- 32/3$ & $32/3$ \\
$\pi (\ln(3))^2 \alpha/\sqrt{3}$ & $0$ & $2/9$ & $0$ & $0$ & $0$ & $- 2/9$ \\
$\pi (\ln(3))^2/\sqrt{3}$ & $2/9$ & $8/9$ & $4/9$ & $2/9$ & $8/9$ & $- 8/9$ \\
\hline
\end{tabular}
\end{center}
\begin{center}
{Table $20$. RI${}^\prime$/SMOM coefficients of $C_F^2$ for two loop $T$ 
amplitudes.}
\end{center}
\end{table}}

\clearpage

{\begin{table}[ht]
\begin{center}
\begin{tabular}{|c||r|r|r|r|r|}
\hline
$a^{(23)}_n$ & $\left. c^{T,(23)}_{(2)\,n} \right|_{\alts}$ & 
$\left. c^{T,(23)}_{(3)\,n} \right|_{\alts}$ & 
$\left. c^{T,(23)}_{(4)\,n} \right|_{\alts}$ & 
$\left. c^{T,(23)}_{(7)\,n} \right|_{\alts}$ & 
$\left. c^{T,(23)}_{(8)\,n} \right|_{\alts}$ \\
\hline
$1$ & $0$ & $- 4/3$ & $- 2/3$ & $- 8/3$ & $0$ \\
$\pi^2 \alpha$ & $- 28/9$ & $- 224/81$ & $- 112/81$ & $- 448/81$ & $28/9$ \\
$\pi^4 \alpha$ & $152/729$ & $- 16/729$ & $- 8/729$ & $- 32/729$ & $- 152/729$ \\
$\zeta(3) \alpha$ & $8/3$ & $0$ & $0$ & $0$ & $- 8/3$ \\
$\Sigma \alpha$ & $2/3$ & $4/3$ & $2/3$ & $8/3$ & $- 2/3$ \\
$\alpha$ & $0$ & $0$ & $0$ & $0$ & $0$ \\
$\pi^2 \alpha^2$ & $8/27$ & $64/81$ & $32/81$ & $128/81$ & $- 8/27$ \\
$\pi^4 \alpha^2$ & $40/729$ & $80/729$ & $40/729$ & $160/729$ & $- 40/729$ \\
$\alpha^2$ & $0$ & $4/3$ & $2/3$ & $8/3$ & $0$ \\
$\pi^2$ & $- 52/3$ & $256/81$ & $128/81$ & $512/81$ & $52/3$ \\
$\pi^4$ & $- 88/243$ & $- 304/729$ & $- 152/729$ & $- 608/729$ & $88/243$ \\
$\zeta(3)$ & $8/3$ & $- 32/3$ & $- 16/3$ & $- 64/3$ & $- 8/3$ \\
$\Sigma$ & $2$ & $- 4/3$ & $- 2/3$ & $- 8/3$ & $- 2$ \\
$s_2(\pi/6)$ & $128$ & $64$ & $32$ & $128$ & $- 128$ \\
$s_2(\pi/6) \alpha$ & $32$ & $0$ & $0$ & $0$ & $- 32$ \\
$s_2(\pi/2)$ & $- 256$ & $- 128$ & $- 64$ & $- 256$ & $256$ \\
$s_2(\pi/2) \alpha$ & $- 64$ & $0$ & $0$ & $0$ & $64$ \\
$s_3(\pi/6)$ & $- 640/3$ & $- 320/3$ & $- 160/3$ & $- 640/3$ & $640/3$ \\
$s_3(\pi/6) \alpha$ & $- 160/3$ & $0$ & $0$ & $0$ & $160/3$ \\
$s_3(\pi/2)$ & $512/3$ & $256/3$ & $128/3$ & $512/3$ & $- 512/3$ \\
$s_3(\pi/2) \alpha$ & $128/3$ & $0$ & $0$ & $0$ & $- 128/3$ \\
$\psi^\prime(1/3)$ & $26$ & $- 128/27$ & $- 64/27$ & $- 256/27$ & $- 26$ \\
$\psi^\prime(1/3) \pi^2 \alpha$ & $- 80/243$ & $160/243$ & $80/243$ & $320/243$ & $80/243$ \\
$\psi^\prime(1/3) \alpha$ & $14/3$ & $112/27$ & $56/27$ & $224/27$ & $- 14/3$ \\
$\psi^\prime(1/3) \pi^2 \alpha^2 $ & $- 40/243$ & $- 80/243$ & $- 40/243$ & $- 160/243$ & $40/243$ \\
$\psi^\prime(1/3) \alpha^2 $ & $- 4/9$ & $- 32/27$ & $- 16/27$ & $- 64/27$ & $4/9$ \\
$\psi^\prime(1/3) \pi^2 $ & $136/81$ & $- 272/243$ & $- 136/243$ & $- 544/243$ & $- 136/81$ \\
$(\psi^\prime(1/3))^2$ & $- 34/27$ & $68/81$ & $34/81$ & $136/81$ & $34/27$ \\
$(\psi^\prime(1/3))^2 \alpha$ & $20/81$ & $- 40/81$ & $- 20/81$ & $- 80/81$ & $- 20/81$ \\
$(\psi^\prime(1/3))^2 \alpha^2$ & $10/81$ & $20/81$ & $10/81$ & $40/81$ & $- 10/81$ \\
$\psi^{\prime\prime\prime}(1/3)$ & $- 2/27$ & $8/27$ & $4/27$ & $16/27$ & $2/27$ \\
$\psi^{\prime\prime\prime}(1/3) \alpha$ & $- 1/27$ & $- 2/27$ & $- 1/27$ & $- 4/27$ & $1/27$ \\
$\pi^3 \alpha/\sqrt{3}$ & $- 58/243$ & $0$ & $0$ & $0$ & $58/243$ \\
$\pi^3/\sqrt{3}$ & $- 232/243$ & $- 116/243$ & $- 58/243$ & $- 232/243$ & $232/243$ \\
$\pi \ln(3) \alpha/\sqrt{3}$ & $- 8/3$ & $0$ & $0$ & $0$ & $8/3$ \\
$\pi \ln(3)/\sqrt{3}$ & $- 32/3$ & $- 16/3$ & $- 8/3$ & $- 32/3$ & $32/3$ \\
$\pi (\ln(3))^2 \alpha/\sqrt{3}$ & $2/9$ & $0$ & $0$ & $0$ & $- 2/9$ \\
$\pi (\ln(3))^2/\sqrt{3}$ & $8/9$ & $4/9$ & $2/9$ & $8/9$ & $- 8/9$ \\
\hline
\end{tabular}
\end{center}
\begin{center}
{Table $21$. Alternative RI${}^\prime$/SMOM scheme coefficients of $C_F^2$ for 
two loop $T$ amplitudes.}
\end{center}
\end{table}}

\end{document}